\documentclass[aps,twocolumn,prl,superscriptaddress,amsmath,showpacs,tightenlines]{revtex4}
\usepackage{epsfig,graphicx,times}
\usepackage{amstext}
\usepackage{amsmath}
\usepackage{amssymb}
\usepackage{graphicx}
\usepackage{latexsym}
\usepackage{bm}
\begin{document}
\title{Correlated emission lasing in harmonic oscillators coupled via a single three-level artificial atom}

\author{Z.H. Peng}\email{zhihui\_peng@riken.jp}
\affiliation{Center for Emergent Matter Science, RIKEN, Wako, Saitama 351-0198, Japan}
\affiliation{Physics Department, Royal Holloway, University of London, Egham, Surrey TW20 0EX, United Kingdom}
\author{Yu-xi Liu}
\affiliation{Institute of Microelectronics, Tsinghua University, Beijing 100084, China}
\affiliation{Tsinghua National Laboratory for Information Science and Technology (TNList), Beijing 100084, China}
\author{J.T. Peltonen}\affiliation{Center for Emergent Matter Science, RIKEN, Wako, Saitama 351-0198, Japan}
\author{T. Yamamoto}\affiliation{NEC Smart Energy Research Laboratories, Tsukuba, Ibaraki 305-8501, Japan}\affiliation{Center for Emergent Matter Science, RIKEN, Wako, Saitama 351-0198, Japan}
\author{J.S. Tsai}\affiliation{Department of Physics, Tokyo University of Science, Kagurazaka, Tokyo 162-8601, Japan}\affiliation{Center for Emergent Matter Science, RIKEN, Wako, Saitama 351-0198, Japan}
\author{O. Astafiev}\email{Oleg.Astafiev@rhul.ac.uk}
\affiliation{Physics Department, Royal Holloway, University of London, Egham, Surrey TW20 0EX, United Kingdom}\affiliation{Center for Emergent Matter Science, RIKEN, Wako, Saitama 351-0198, Japan}\affiliation{National Physical Laboratory, Teddington, TW11 0LW, United Kingdom}\affiliation{Moscow Institute of Physics and Technology, Dolgoprudny, 141700, Russia}

\begin{abstract}
%A single superconducting artificial atom provides a unique basis for coupling electromagnetic fields and photons which is hardly achieved with a natural atom. Bringing a pair of harmonic oscillators into resonance with the transitions of a three-level atom converts atomic spontaneous processes into correlated emission dynamics. We present experimental demonstration of two-mode correlated emission lasing in harmonic oscillators coupled via a fully controllable three-level artificial atom. This is fundamentally different from the typical systems of quantum optics, where atoms are coupled through a linear oscillator. The correlation of emissions with two different colors reveals itself as equally narrowed linewiths and quenching of their mutual phase diffusion. The mutual linewidth is more than four orders of magnitude narrower than the Schawlow-Townes limit. The interference between the different color lasing fields demonstrates that the two-mode fields are strongly correlated.
A single superconducting artificial atom can be used for coupling electromagnetic fields up to the single photon level due to easily achieved strong coupling regime. Bringing a pair of harmonic oscillators into resonance with the transitions of a three-level atom converts atomic spontaneous processes into correlated emission dynamics. We present experimental demonstration of two-mode correlated emission lasing in harmonic oscillators coupled via a fully controllable three-level superconducting quantum system (artificial atom). The correlation of emissions with two different colors reveals itself as equally narrowed linewiths and quenching of their mutual phase diffusion. The mutual linewidth is more than four orders of magnitude narrower than the Schawlow-Townes limit. The interference between the different color lasing fields demonstrates that the two-mode fields are strongly correlated.
\end{abstract}
%\date{\today}
\maketitle


Excited free atoms relax to their ground states via incoherent spontaneous emission processes \cite{Scully1997}. In an ensemble of three-level atoms -- the fundamental objects of quantum optics -- coupling of the lowest pair of levels to a harmonic oscillator leads to conventional lasing if incoherent relaxation from the second excited state to the first excited state creates population inversion.
The situation is dramatically changed when both transitions of the atoms are coupled to independent oscillators: the spontaneous emission processes are replaced by coherent energy exchange and correlated dynamics between the oscillators, known as correlated emission lasing (CEL) \cite{Scully1997,Scully1985,Scully1988,Winters1990,Steiner1995,Abich2000}.
The correlated dynamics reveals itself as a reduction in relative random phase-diffusion noise, resulting in the suppression of mutual peak width below the Schawlow-Townes limit \cite{ST}.
The suppression of phase-diffusion noise has been experimentally observed for two polarization modes in a HeNe laser~\cite{Winters1990,Steiner1995,Abich2000}. However, the best result observed so far has been about $2.3\%$ of the Schawlow-Townes limit for an ensemble of natural atoms~\cite{Abich2000}. Recently, the classical lasing effect has been reproduced on single atoms \cite{McKeever2003} or single superconducting quantum systems (artificial atoms) ~\cite{Astafiev2007,Grajcar2008,Oelsner2013}, in which strong coupling to circuit elements can be easily achieved~\cite{Astafiev2007,Grajcar2008,Oelsner2013,Wallraff2004, Astafiev2010}. In this Letter, we demonstrate the coherent dynamics of two harmonic modes with different frequencies in a transmission line resonator (TLR) coupled through a single three-level artificial atom. Oppositely to the ensemble of atoms with uncorrelated dephasing in different ones, the single atom allows to ideally suppress the phase diffusion. We demonstrate quenching of the mutual phase diffusion to the level better than $10^{-4}$. The correlation between two lasing fields is displayed in the interference between them.

In addition, we point out that our system presents a fundamentally new circuit. In usual quantum electrodynamics, atoms are coupled through a resonator mode, however the reversed circuit with linear oscillators coupled via transitions of a quantum system is very difficult, if not impossible, to realize with natural atoms, spins or quantum dots. Such a system presents a novel approach and allows to demonstrate a series of qualitatively new phenomena of quantum optics.

% ----------------------FIGURE1 -----------------------------------
\begin{figure}
\center
\includegraphics[scale=0.4]{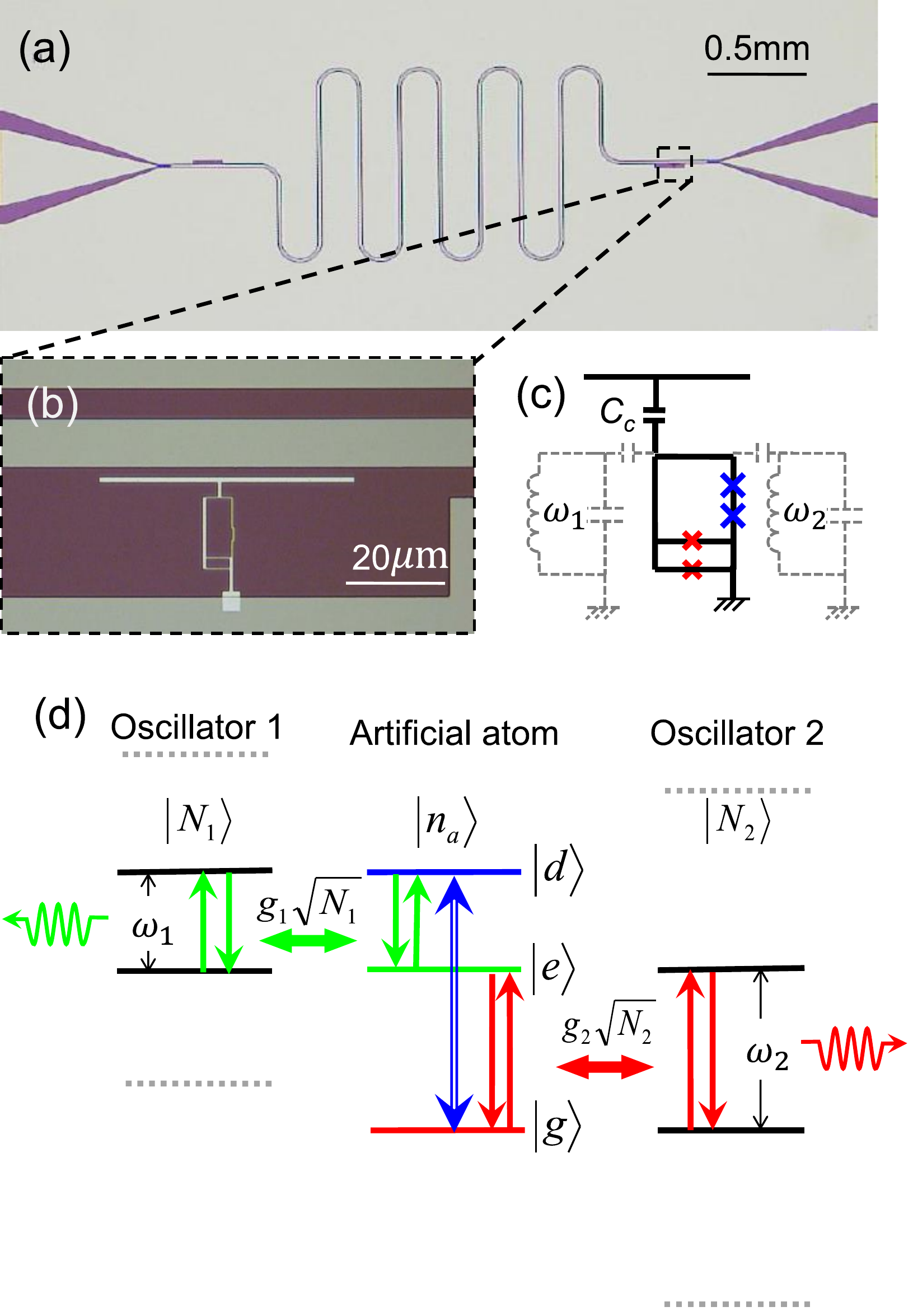}
\caption{{Device description.} ({a}) Optical micrograph of the device. The meandering structure is a TLR made of a 50-nm-thick Nb film capacitively coupled to external coplanar waveguides.
({b}) Magnified micrograph of an artificial atom based on the tunable gap superconducting flux qubit geometry. The $\alpha$-loop (with red junctions) allows the flux tunnelling energy to be tuned.
({c}) Simplified circuit model of the atom capacitively coupled to the TLR through capacitance $C_c$. The atom is effectively coupled to two oscillators with the frequencies $\omega_1$ and $\omega_2$, schematically shown by the dashed lines.
({d}) Principles of the device operation. The transitions $|d\rangle$-$|e\rangle$ and $|e\rangle$-$|g\rangle$ are resonant with the first and second oscillators, respectively. The transition $|g\rangle$-$|d\rangle$ is driven by a classical microwave pumping field. }
\label{picture}
\end{figure}
%% ----------------------FIGURE -----------------------------------

As shown in Fig.~\ref{picture}(a), Fig.~\ref{picture}(b) and Fig.~\ref{picture}(c), the single three-level artificial atom with cyclic transitions~\cite{Liu2005}  is based on a \textquotedblleft tunable gap flux qubit\textquotedblright~circuit~\cite{Paauw2009,Zhu2010} capacitively coupled to a multimode transmission line resonator. Such a coupling for the flux qubits has been realised in \cite{Inomata2012} and the transition matrix elements for the three-level system calculated in \cite{Supp} are consistent with our experiment. The artificial atom is situated in the voltage antinode of the resonator.
The transition frequencies ($\omega_{eg}$, $\omega_{dg}$ and $\omega_{de}$) between the three lowest levels of the atom $|g\rangle$, $|e\rangle$ and $|d\rangle$ (the ground, first excited and second excited states) are controlled by an external magnetic flux, reaching minimal $\omega_{eg}$ and $\omega_{dg}$ at half-integer flux quanta $\Phi_N = (N+\frac{1}{2})\Phi_{0}$ (where $N$ is an integer).
Compared with systems based on the conventional flux qubit geometry \cite{Mooij1999,Wal2000}, our atom has additional tunability owing to the implementation of an $\alpha$-loop -- a dc SQUID, which allows the flux tunnelling energy $\Delta$ to be changed by choosing $\Phi_N$. By selecting $\Phi_N$ and additionally tuning $\delta\Phi$  ($\ll \Phi_0$), one can adjust the transition frequencies $\omega_{eg}$ and $\omega_{de}$ to the resonance with the two lowest modes of the resonator -- the two independent fixed-frequency oscillators $r_1$ and $r_2$ with the frequencies $\omega_1$ and $\omega_2$, respectively.

The entire system presented in Fig.~\ref{picture}(d) can be described by the state $|N_{1} n N_{2}\rangle=|N_1\rangle\otimes|n\rangle\otimes|N_{2}\rangle$ with the three quantum numbers $n$, $N_1$ and $N_2$, denoting the three-level atomic states $|n\rangle=\lbrace |g\rangle, |e\rangle,|d\rangle\rbrace$ and the Fock states of resonators $|N_{1,2}\rangle=\lbrace|0\rangle,|1\rangle,|2\rangle,...\rbrace$ with the photon occupation $N_{1,2}$ in the first and second oscillators.
The full Hamiltonian of the two harmonic oscillators coupled via transitions of the three-level atom and driven by a microwave with the frequency $\omega_p \approx \omega_{dg}$ and amplitude $2\hbar\Omega$ is
\begin{eqnarray} \label{effectHami}
H &=&\hbar \omega_{1} a_{1}^{\dag}a_{1}+\hbar\omega_{2}a_{2}^{\dag}a_{2}+\hbar\omega_{dg} \sigma_{dd}+\hbar \omega_{eg}\sigma _{ee} \\  \nonumber
&&+\hbar g_{1}\left( a_{1}^{\dag}\sigma _{ed}+a_{1}\sigma _{de}\right)+\hbar g_{2}\left( a_{2}^{\dag }\sigma _{ge}+a_{2}\sigma_{eg}\right)\\ \nonumber
&&+2\hbar \Omega(\sigma_{dg}+\sigma_{gd}) \cos\omega_p t.
\end{eqnarray}
Here, $\sigma_{jk}=|j\rangle\langle k|$ is the projection/transition operator for the atomic states with $\{j,k\}=\{e,g,d\}$. $a^\dag_1$ ($a_1$) and $a^\dag_2$ ($a_2$) are the creation (annihilation) operators of the first and second oscillators, respectively. The atomic states $|g\rangle$ and $|d\rangle$ are coupled by the pumping field. To eliminate the time dependence, the Hamiltonian can be further simplified using the rotating wave approximation, in which the two oscillators appear to be resonantly coupled via the quantum system. The dissipative dynamics of this artificial atom-resonator coupled system is described by the Markovian master equation with accounting energy relaxation and decoherence in the system \cite{Supp}.

\begin{figure*}
\center
\includegraphics[scale=0.9]{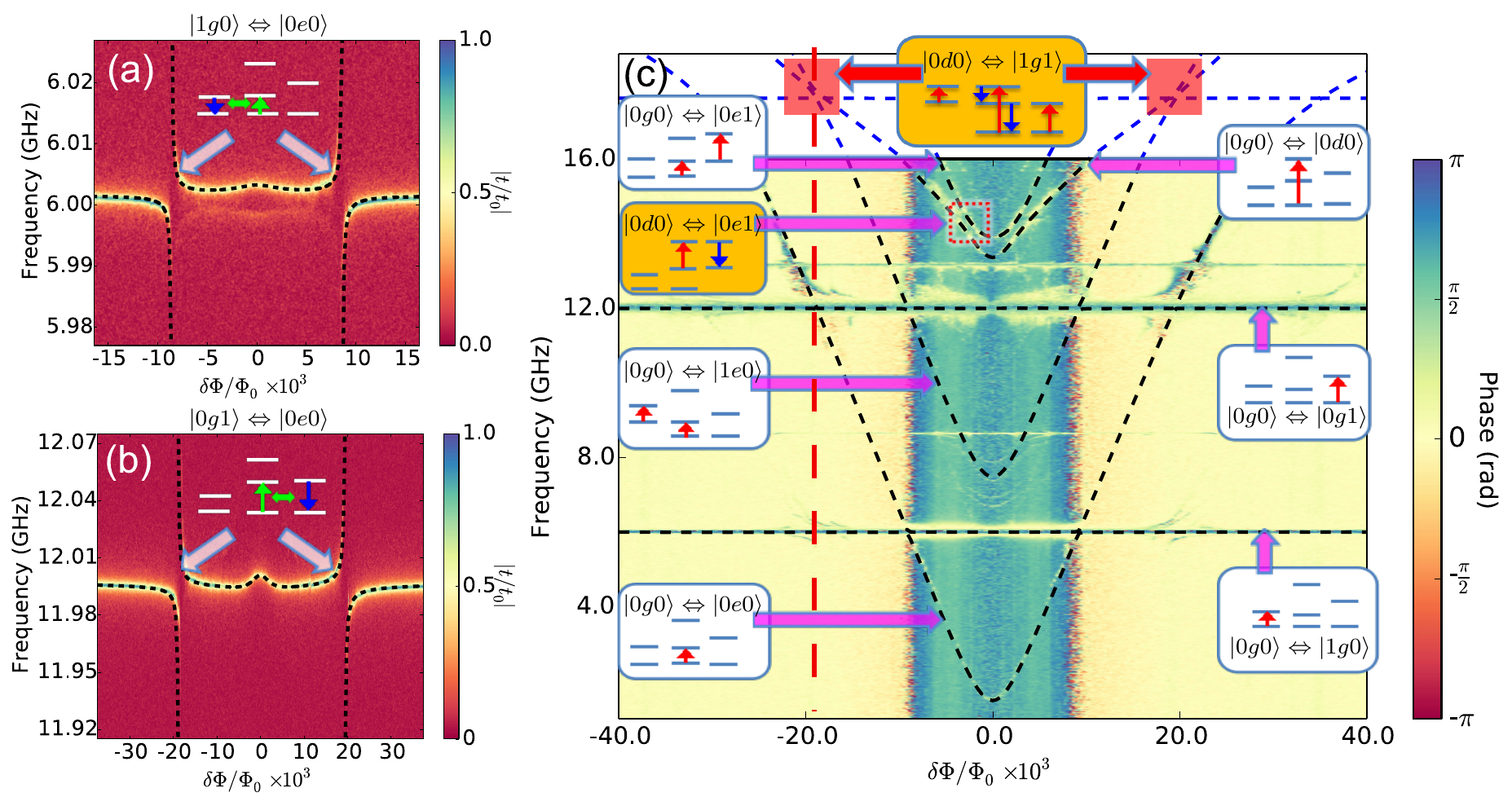}
\caption{{Energy levels of the three-level atom-resonator coupled system.} {(a) and (b)} Normalized transmission amplitude through the resonator modes as a function of $\delta\Phi$. The black dashed lines are theoretical fittings from the full system Hamiltonian. ({c}) Intensity plot of the phase of a transmission signal as a function of driving frequency and $\delta\Phi$. The black dashed lines represent theoretical calculations. Our goal is to study the double resonance shown in Fig. 1(d), which takes place under the conditions marked by the red squares.}
\label{Spectroscopy}
\end{figure*}

We perform microwave characterization (transmission and emission measurements) of the device in  a dilution refrigerator at a temperature of 30 mK. The device fabrication process and experimental setup are described in detail in Supplementary Materials \cite{Supp}.
From the transmission measurement, $\omega_{1}/2\pi\approx6.0016\,$GHz and $\omega_{2}/2\pi\approx11.9979\,$GHz with the corresponding linewidths (decay rates) in the oscillators of $\kappa_{1}/2\pi=0.63\,$MHz and  $\kappa_{2}/2\pi=1.94\,$MHz at the base temperature.
We find that $\Phi_N\approx1.5\Phi_{0}$ ( $\Delta = h\times1.51\,$GHz) and $\delta\Phi \sim 18\times 10^{-3}\Phi_0$ is one of the best choices to reach the desired double resonance. To characterise the atom-resonator interaction, we measure the normalized transmission through the system $|t/t_0|$ around $\omega_{1}$ and $\omega_{2}$ as a function of the probing frequency and flux bias $\delta\Phi$ using a vector network analyser (Fig.~\ref{Spectroscopy}(a) and Fig.~\ref{Spectroscopy}(b)), where $t_0$ is the maximal transmission amplitude.
Anticrossings due to vacuum Rabi splittings are clearly observed when $\omega_{eg}\approx\omega_{1}$ in Fig.~\ref{Spectroscopy}(a) and $\omega_{eg}\approx\omega_{2}$ in Fig.~\ref{Spectroscopy}(b), respectively. The coupling strengths of $g_{3}=2\pi \times36\,$MHz (corresponding to the interaction $|0 e 0\rangle <=> |1 g 0\rangle$ at $\delta\Phi \approx \pm 9\times 10^{-3}\Phi_0 $) and $g_{2}=2\pi \times78\,$MHz ($|0 e 0\rangle <=> |0 g 1\rangle$  at $\delta\Phi \approx \pm 19.5\times10^{-3}\Phi_0 $) are obtained by fitting the anticrossings.

% ----------------------FIGURE2 -----------------------------------
% ----------------------FIGURE -----------------------------------

In Fig.~\ref{Spectroscopy}(c), we use the two-tone microwave spectroscopy technique to find the energy levels of the atom-resonator coupled system \cite{Abdumalikov2008}. We measure the transmission at a fixed frequency close to $\omega_1$ and sweep the pumping frequency in the range from 1 to 16 GHz (limited by the bandwidth of the entire system). However, we show our theoretical calculations (which are in a good agreement with the experimentally measured lines) above the bandwidth limit by dashed lines. The applied probing power is sufficiently low to keep the average number of photons inside the resonator less than one ($\langle N_1 \rangle < 1$).
At $\delta\Phi \neq 0$, all the transitions between the three levels are allowed~\cite{Liu2005}.
The black dashed lines in Fig.~\ref{Spectroscopy}(a), Fig.~\ref{Spectroscopy}(b) and Fig.~\ref{Spectroscopy}(c) are the fitting curves obtained from the Hamiltonian of the atom-resonator system \cite{Inomata2012}. The dashed red box in Fig.~\ref{Spectroscopy}(c) shows the avoided crossing due to the interaction between $|0 d 0\rangle$ and $|0 e 1\rangle$ characterised by $g_{4}=2\pi \times205\,$MHz at $\delta\Phi \approx \pm 2.5\times10^{-3}\Phi_0$, where the second excited atomic state is converted into a photon in $r_2$ and the first excited atomic state. Note that $g_4$ extracted from the experiment is also consistent with the theoretical calculations~\cite{Supp}.

\begin{figure}
\center
\includegraphics[scale=0.3]{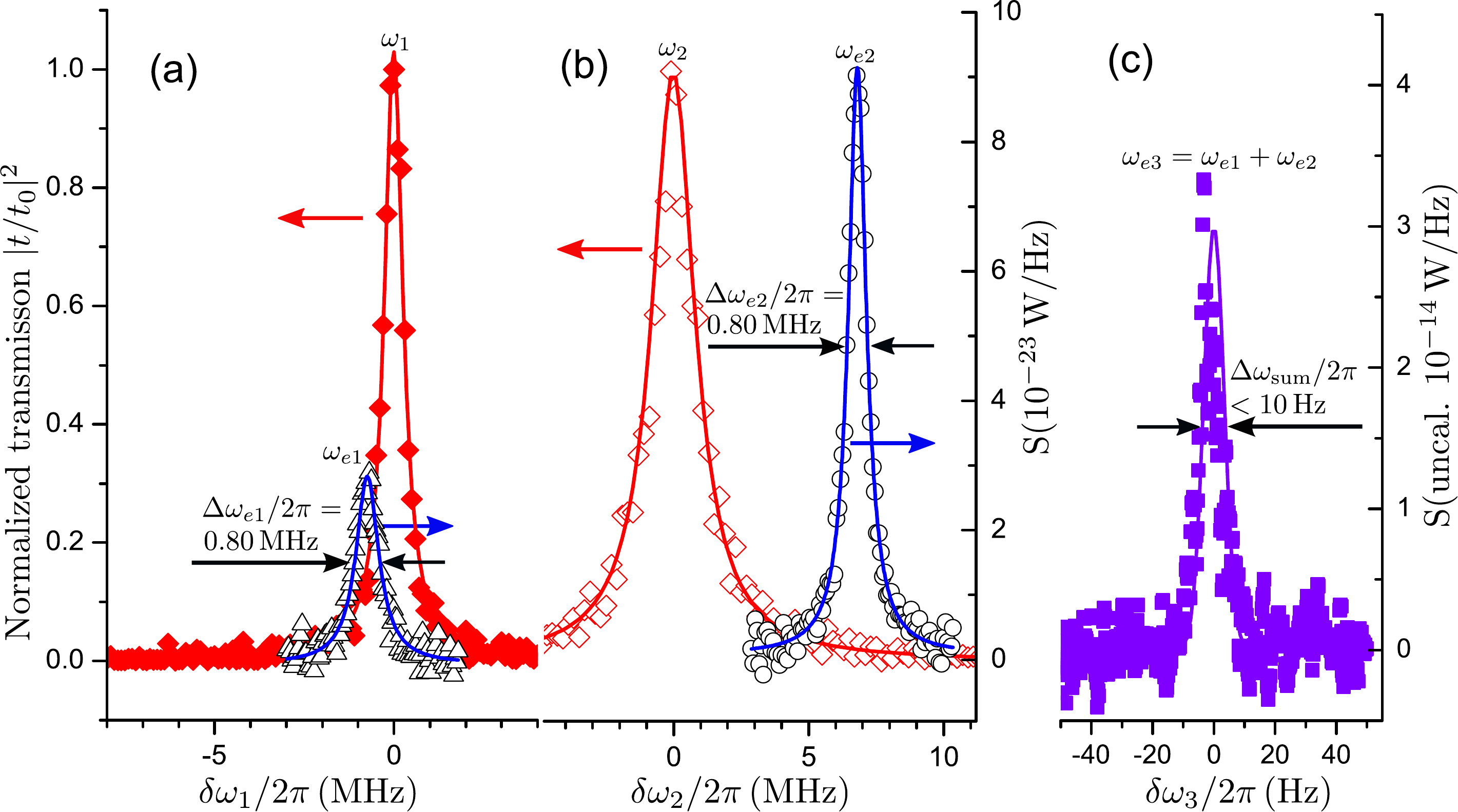}
\caption{{Emission spectra of two-mode CEL and quenching of the phase diffusion noise. }(a) and (b) Normalized transmission through oscillators 1 and 2 (red closed and open rhombus) and emission spectra from the oscillators (black open triangulars and circles). The emission spectra have the same widths of $\Delta\omega_{e1}=\Delta\omega_{e2}=2\pi\times0.80$ MHz. (c) Quenching of the mutual phase diffusion. The two frequency emissions are amplified and mixed. The resulting upconverted signal (uncalibrated) is very narrow and limited by the SA bandwidth.}
\label{Emission}
\end{figure}

To couple the oscillators via the atomic transitions, we set the bias to $\delta\Phi_b = -18\times10^{-3}\Phi_0$ (red dashed line), where $\omega_{eg} \approx \omega_2$ and $\omega_{de} \approx \omega_1$, and apply the external microwave drive at a pumping frequency of $\omega_p = 2\pi\times18.0055\,$GHz. Although $g_1$ cannot be directly measured, we take its value as 2$\pi\times90$~MHz from simulations, which well reproduces other couplings (detailed in Supplementary Materials \cite{Supp}). Strong emission peaks appear when the pumping power exceeds $P \approx -100$ dBm. The emission disappears, when the atom is detuned from the double resonance. We assume that the threshold has a different nature from conventional lasing and is due to the remaining detuning between $\omega_{de}$, $\omega_1$ and/or $\omega_{eg}$, $\omega_2$, compensated by Rabi splitting, which according to our estimates is $\Omega/2\pi \approx$ 900~MHz at $\omega_{dg}$. This is supported by our simulations~\cite{Supp}. Note also that we expect another resonance close (but not equal) to $\omega_p$. However it does not affect the overall picture because the drive is essentially classical with many photons in the mode. The emission spectra of the coupled system in the vicinity of $\omega_1$ and $\omega_2$ are simultaneously monitored by two spectral analysers (SAs). We use the techniques developed in \cite{Abdumalikov2011} to eliminate the noise background of the low temperature amplifier by subtracting traces of OFF from ON emission spectra. The results are shown in Fig.~\ref{Emission}(a) (black open triangles) and Fig.~\ref{Emission}(b) (black open circles). Importantly, the peaks are well fitted by equal width Lorentzian curves with $\Delta\omega_{e1} \approx \Delta\omega_{e2} = 2\pi\times0.80\,$MHz, indicating the interaction between the oscillators. Note that $\Delta\omega_{e1,2}$ is about two times lower than $\kappa_2$ and three times lower than the effective relaxation rate of the entire system of $\kappa_1 + \kappa_2 \approx 2\pi\times 2.57$ MHz. The center frequencies of the two emission spectra of $\omega_{e1} = 2\pi\times$6.00086~GHz and $\omega_{e2} = 2\pi\times$~12.00471~GHz are somewhat shifted from $\omega_1$ and $\omega_2$ by $\omega_{e1} - \omega_{1}=- 2\pi\times0.74\,$MHz and $\omega_{e2} - \omega_{2}=2\pi\times 6.81\,$MHz, which is probably due to dispersive shifts from the detuning between the artificial atom and the resonator modes \cite{Inomata2012}.
We emphasise that $\omega_{e1} + \omega_{e2} = \omega_p$ holds with high accuracy. Although the emission frequencies slightly depend on other parameters, such as pumping power and frequency, the above equation still holds~\cite{Supp}.

The estimated total emission powers, obtained from the extracted area of the emission curves, are $P_{1}\approx-134.4$ dBm and $P_{2}\approx-129.4$ dBm, with an accuracy of about 3 dBm due to the uncertainty in the calibration of our setup. They  roughly correspond to $\langle N_{1}\rangle \approx 5$ photons and $\langle N_{2}\rangle \approx 2$ photons at $r_1$ and $r_2$, respectively. (The photon number is derived using the equation $\langle N_k\rangle = 2 P_k / (\hbar\omega_k\kappa_k)$, where $k = \{1,2\}$ and the prefactor 2 is due to the escape of photons from either end of the resonator with equal probability.) The ratio $\langle N_1\rangle/\langle N_2\rangle$ is found to be close to the expected value as the photon creation rate in each mode under stationary conditions is expected to be given by $\langle N_1\rangle \kappa_1 = \langle N_2\rangle \kappa_2$, which originates from the energy conservation law: each absorbed photon at $\omega_p$ is split into a pair of photons at $\omega_{e1}$ and $\omega_{e2}$.

% ----------------------FIGURE3 -----------------------------------
%% ----------------------FIGURE -----------------------------------
%----------------------FIGURE4 -----------------------------------
\begin{figure}
\center
\includegraphics[scale=0.21]{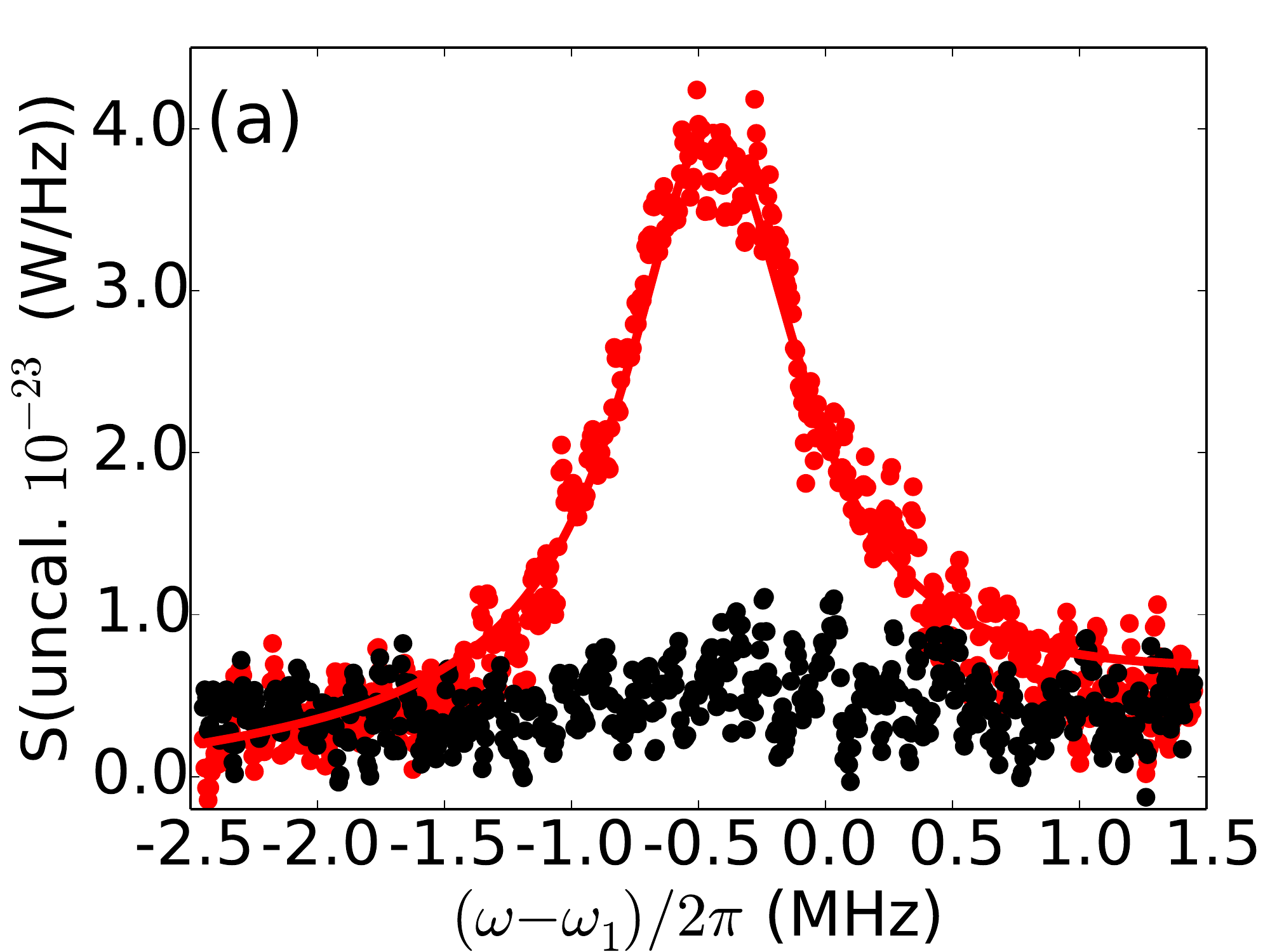}
\includegraphics[scale=0.21]{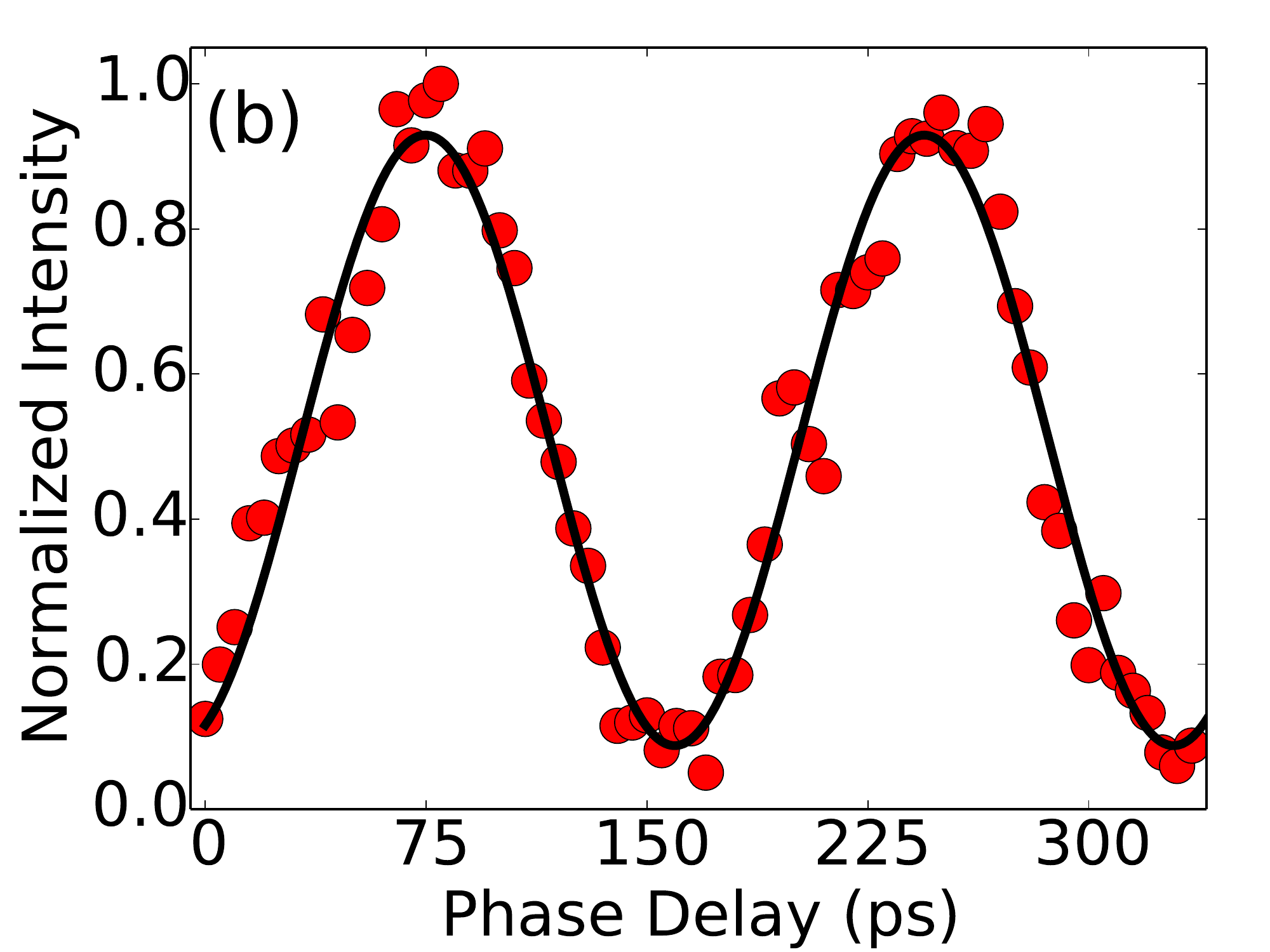}
\caption{{Correlation between the two emission modes. }(a) Constructive (red dots) and destructive (black dots) interference peaks of the emissions with two different colors. To observe the interference, the $\omega_{e2}$-signal is downconverted to $\omega_{p}-\omega_{e2}$ and a variable delay is introduced in the $\omega_{e1}$ emission signal before summing it with $\omega_p - \omega_{e2}$. (b) Interference fringes of the peak amplitudes from {{(a)}} as a function of the phase delay between the two signals (red dots). The amplitudes of both signals are adjusted to obtain nearly maximal modulation. The intensity derived from the experimental peaks (red dots) is fitted by a sine function (black curve). %(c) Time dependent of $(\Delta\hat{u})^2+(\Delta\hat{v})^2$ for initial vacuum states in terms of the normalized time $g_2t$.
}
\label{Correlation}
\end{figure}
% ----------------------FIGURE -----------------------------------

%(\textit{OA: Discussion. Should be corrected. Could you work on that?}) Below are several conclusions supported by the theory presented in [Supplementary]. The obtained results are in a good agreement with our theoretical analysis. The correlated dynamics appears through the equal width of the peak and quench of the mutual phase diffusion between the two emissions.

To demonstrate quenching of the phase diffusion in the two-mode CEL, we mix the amplified emission signals at $\omega_{e1}$ and $\omega_{e2}$ using a conventional microwave mixer. Then, we measure the signal around the sum of the frequencies
$\omega_{e1} + \omega_{e2}$ by the SA~\cite{Supp}. Note that the phase diffusion in our case is different from the relative phase diffusion in Refs.~\cite{Winters1990,Steiner1995,Abich2000} and we call it mutual phase diffusion. To avoid direct leakage of the pumping tone through our circuit, we filter out high frequencies before mixing and confirm the absence of any outcoming signal at $\omega_p$.
Next, we simultaneously monitor the signals at $\omega_{e2}$ ($\omega_{e1}$) and $\omega_{e1} + \omega_{e2}$. The result shown in Fig.~\ref{Emission}(c) demonstrates a strong narrow sum signal at a frequency of exactly $\omega_p$ only when emissions at $\omega_{e1,2}$ occur. Despite mixing a pair of signals of $\Delta\omega_e$ width, the resulting peak width of $\Delta\omega_{\rm{sum}} = 2\pi\times9.4\,$Hz is found to be limited by the bandwidth of the SA ($10\,$Hz). We note that the Schawlow-Townes condition for the linewidth is expected to be modified in our system to $\Delta\omega_{\rm{ST}} = (\kappa_1 + \kappa_2)/(2 N_{\rm{tot}})$, which is about $250\,$kHz for 10 photons. The experimentally measured linewidths of different color emissions satisfy the condition $\kappa_1 + \kappa_2 > \Delta \omega_{e1,2} > \Delta\omega_{\rm{ST}}$, which usually takes place for conventional lasers due to additional broadening mechanisms.
However, the mutual phase diffusion noise (characterised by $\Delta\omega_{\rm{sum}}$) is quenched being at least four orders of magnitude lower than $\Delta\omega_{\rm{ST}}$, which can be explained by the theoretical model~\cite{Supp}.

To further quantify the correlation between the emissions, we downconvert the emission at $\omega_{e2}$ by mixing it with the $\omega_p$ tone and select out the difference frequency $\omega_p - \omega_{e2}$. Then, we linearly add the resulting signal to that at $\omega_{e1}$ using a linear adder and measure the resulting signal by the SA~\cite{Supp}. The intensity of the observed signal is extracted by fitting with a Lorentzian curve (red curve in Fig.~\ref{Correlation}(a)) and then is studied as a function of the phase delay introduced in the $\omega_{e1}$ emitted signal. As shown in Fig.~\ref{Correlation}(b), we observe the sinusoidal oscillations of the intensity of the interfering signal as a function of the delay, indicating a strong correlation between the two lasing fields. %The quantum correlation signal is easily destroyed by the high noise temeprature commercial mixer we used at room temperature. It is difficult for us observe the experimental evidence of quantum correlation between two lasing fields: the noise floor of destructive interference peak should be lower than the consctructive interference peak in Fig.~\ref{Correlation}(a). However, as shown in Fig.~\ref{Correlation}(c),
The interference is an additional evidence for the quenching of the mutual phase diffusion.

Finally, we briefly summarise other aspects of the system, consisting of the two oscillators coupled through the transitions of the three-level atom, which can be verified experimentally in the future. The single three-level atom CEL system provides an ideal testbed for studying the nonlinear coupling of light in cavity modes.
It was theoretically found that in contrast to two-mode squeezed states by parametric amplifiers~\cite{Milburn1981,Ou1992,Eichler2011,Bergeal2012,Flurin2012}, a bright and arbitrarily high degree squeezed lights can be generated inside a cavity by a system of atoms interacting with a two-mode cavity because of the strong atomic nonlinearity~\cite{Scully1988,Gea1990a,Gea1990b}. Moreover, the nonlinear coupling of light in the two cavity modes can lead to two-mode squeezed states~\cite{Scully1997}. We calculate the time dependence of the sum of the quantum fluctuations of the two lasing fields $(\Delta\hat{u})^2+(\Delta\hat{v})^2$ based on the experimental parameters and find that it can be smaller than two \cite{Xiong2005,Tan2005,Qamar2007,Duan2000}, which suggests the possible existence of quantum correlation between the two fields in our sample for the short time after tuning on the pump (details are given in Supplementary Materials \cite{Supp}). In addition, the quantum coupler provides an ideal (the strongest possible) nonlinearity for weak fields up to the single quantum (photon) level in the strong coupling regime. For example, there is a theoretical discussion about strong coupling of a phonon mode and a photon mode with a single superconducting three-level artificial atom at single-photon level \cite{Xue2015}.

In conclusion, we have demonstrated correlated dynamics between two oscillators coupled via transitions of a three-level macroscopic artificial atom.
The dynamics is observed as the quantum noise quenching of the mutual phase diffusion
in the correlated emission lasing. Our results are the first step to generate the quantum correlation and entanglement between two lasing fields with a single atom, which is an important step towards constructing a quantum network in the microwave domain. This work may also lead to an improvement in ultrahigh-sensitive interferometric measurements\cite{Scully1997}.

\begin{acknowledgements}
Z.H. Peng would like to thank F. Yoshihara and K. Kusuyama for useful help on sample fabrication, Y. Kitagawa for preparing Nb wafers, and J.Q. Liao and S. Qamar for useful discussion. This work was supported by MEXT kakenhi \textquotedblleft Quantum Cybernetics\textquotedblright~and the JSPS through its FIRST Program. This work was funded by ImPACT Progam of Council for Science, Technology and Innovation (Cabinet Office, Government of Japan). This work was done within the project EXL03 MICROPHOTON of the European Metrology Research Programme (EMRP). The EMRP is jointly funded by the EMRP participating countries within EURAMET and the European Union. Y.X. Liu is supported by NSFC under Grant Nos. 61025022, 91321208 and National Basic Research Program of China Grant No. 2014CB921401.
\end{acknowledgements}

\end{document}

% Place the author information here.  Please hand-code the contact
% information and notecalls; do *not* use \footnote commands.  Let the
% author contact information appear immediately below the author names
% as shown.  We would also prefer that you don't change the type-size
% settings shown here.

\title{Supplementary Materials for \textquotedblleft Correlated emission lasing in harmonic oscillators coupled via a single three-level artificial atom\textquotedblright}

\author{Z.H. Peng}\email{zhihui\_peng@riken.jp}
\affiliation{Center for Emergent Matter Science, RIKEN, Wako, Saitama 351-0198, Japan}
\affiliation{Physics Department, Royal Holloway, University of London, Egham, Surrey TW20 0EX, United Kingdom}
\author{Yu-xi Liu}
\affiliation{Institute of Microelectronics, Tsinghua University, Beijing 100084, China}
\affiliation{Tsinghua National Laboratory for Information Science and Technology (TNList), Beijing 100084, China}
\author{J.T. Peltonen}\affiliation{Center for Emergent Matter Science, RIKEN, Wako, Saitama 351-0198, Japan}
\author{T. Yamamoto}\affiliation{NEC Smart Energy Research Laboratories, Tsukuba, Ibaraki 305-8501, Japan}\affiliation{Center for Emergent Matter Science, RIKEN, Wako, Saitama 351-0198, Japan}
\author{J.S. Tsai}\affiliation{Department of Physics, Tokyo University of Science, Kagurazaka, Tokyo 162-8601, Japan}\affiliation{Center for Emergent Matter Science, RIKEN, Wako, Saitama 351-0198, Japan}
\author{O. Astafiev}\email{Oleg.Astafiev@rhul.ac.uk}
\affiliation{Physics Department, Royal Holloway, University of London, Egham, Surrey TW20 0EX, United Kingdom}\affiliation{Center for Emergent Matter Science, RIKEN, Wako, Saitama 351-0198, Japan}\affiliation{National Physical Laboratory, Teddington, TW11 0LW, United Kingdom}\affiliation{Moscow Institute of Physics and Technology, Dolgoprudny, 141700, Russia}

%%%%%%%%%%%%%%%%% END OF PREAMBLE %%%%%%%%%%%%%%%%

\maketitle

\section{Two-mode CEL theory}
\subsection{Master equation of reduced density matrix for two cavity modes}
In the rotating reference frame with $U=\exp(-i\omega_p\sigma_{dd}t-i\omega_{p}a_{1}^{+}a_{1}t)$,
the Hamiltonian for the
three-level system interacting with two cavity fields (Eq.~(1)) is simplified to
\begin{eqnarray}
H&=&\hbar\Delta_{1}a^{\dagger}_{1}a_{1}+\hbar\omega_{2}a^{\dagger}_{2}a_{2}+%
\hbar\omega_{eg}\sigma_{ee}+\hbar\Delta_{2}\sigma_{dd}  \notag \\
&+&\hbar[\Omega\exp(i\phi)\sigma_{gd}
+g_{1}a^{\dagger}_{1}\sigma_{ed}+g_{2}a^{\dagger}_{2}\sigma_{ge}]+\text{H.c.}
\end{eqnarray}
with two detunings, $\Delta_{1}=\omega_{1}-\omega_{p}$ and $%
\Delta_{2}=\omega_{dg}-\omega_{p}$. Taking the reduced density
matrix of the three-level system and two cavity fields, $\rho$, we
can write the master equation
\begin{eqnarray*}  \label{Lindblad}
\frac{\partial \rho}{\partial t} &=&-\frac{i}{\hbar}[H,\rho]+\frac{\gamma
_{31}}{2}\left[ 2\sigma _{gd}\rho \sigma _{dg}-\sigma _{dg}\sigma _{gd}\rho
-\rho \sigma _{dg}\sigma _{gd}\right] \\
&&+\frac{\gamma _{32}}{2}\left[ 2\sigma _{ed}\rho \sigma _{de}-\sigma
_{de}\sigma _{ed}\rho -\rho \sigma _{de}\sigma _{ed}\right] \\
&&+\frac{\gamma _{21}}{2}\left[ 2\sigma _{ge}\rho \sigma _{eg}-\sigma
_{eg}\sigma _{ge}\rho -\rho \sigma _{eg}\sigma _{ge}\right] \\
&&+\frac{\gamma _{22}}{2}\left[ 2\sigma _{ee}\rho \sigma _{ee}-\sigma
_{ee}\rho -\rho \sigma _{ee}\right] \\
&&+\frac{\gamma _{33}}{2}\left[ 2\sigma _{dd}\rho \sigma _{dd}-\sigma
_{dd}\rho -\rho \sigma _{dd}\right] \\
&&+\kappa _{1}\left( 2a_{1}\rho a_{1}^{\dag }-a_{1}^{\dag }a_{1}\rho -\rho
a_{1}^{\dag }a_{1}\right) \\
&&+\kappa _{2}\left( 2a_{2}\rho a_{2}^{\dag }-a_{2}^{\dag }a_{2}\rho -\rho
a_{2}^{\dag }a_{2}\right),
\end{eqnarray*}%
with the decay rates $\kappa_{1}$ and $\kappa_{2}$ of the two cavity modes and $\gamma_{jk}$ of the atom, where $\{k,j\}={1,2,3}$.
Using standard methods of laser theory as in Ref.~\cite{Scully-book}, we can
eliminate the variables of the three-level system and keep to the first
order of the coupling constants $g_{i}$ ($i=1,2$). The following is the
master equation for the reduced density matrix of the two modes of the
cavity field
\begin{eqnarray}  \label{eq:7}
\frac{\partial \rho_{f}}{\partial t}&=&-i[\Delta_{1}a^{\dagger}_{1}a_{1}+%
\omega_{2}a^{\dagger}_{2}a_{2},\rho_{f}] \\
&+&\left\{\left[\alpha_{1}\rho^{(0)}_{gg,A}+\mu^{*}_{1}\rho_{dg,A}^{(0)}%
\right](a_{2}\rho_{f}a^{\dagger}_{2}-\rho_{f}a^{\dagger}_{2}a_{2})+\text{H.c.%
}\right\}  \notag \\
&+&\left\{\left[\alpha_{2}\rho^{(0)}_{dd,A}+\mu^{*}_{2}\rho_{gd,A}^{(0)}%
\right](a^{\dagger}_{1}\rho_{f}a_{1}-\rho_{f}a_{1}a_{1}^{\dagger})+\text{H.c.%
}\right\}  \notag \\
&+&\left\{\alpha_{1}\rho_{ee,A}^{(0)}(a^{\dagger}_{2}\rho_{f}a_{2}-a_{2}a^{%
\dagger}_{2}\rho_{f})+\text{H.c.}\right\}  \notag \\
&+&\left\{\alpha_{2}\rho_{ee,A}^{(0)}(a_{1}\rho_{f}a^{\dagger}_{1}-a^{%
\dagger}_{1}a_{1}\rho_{f})+\text{H.c.}\right\}  \notag \\
&+&\left\{\left(\beta_{1}\rho_{gd,A}^{(0)}+\nu^{*}_{2}\rho_{dd,A}^{(0)}%
\right)(a_{2}\rho_{f}a_{1}-\rho_{f}a_{1}a_{2})+\text{H.c.}\right\}  \notag \\
&+&\left\{\left(\beta_{2}\rho_{gd,A}^{(0)}+\nu_{1}\rho_{gg,A}^{(0)}%
\right)(a_{2}\rho_{f}a_{1}-a_{1}a_{2}\rho_{f})+\text{H.c.}\right\}  \notag \\
&+&\left\{\nu_{2}\rho_{ee,A}^{(0)}(a^{\dagger}_{2}\rho_{f}a^{\dagger}_{1}-%
\rho_{f}a^{\dagger}_{1}a^{\dagger}_{2})+\text{H.c.}\right\}  \notag \\
&+&\left\{\nu_{1}\rho_{ee,A}^{(0)}(a_{1}\rho_{f}a_{2}-\rho_{f}a_{1}a_{2})+%
\text{H.c.}\right\}  \notag \\
&+&\sum_{i=1}^{2}\kappa_{i}(2a_{i}\rho_{f}a_{i}^{\dagger}-a^{%
\dagger}_{i}a_{i}\rho_{f}- \rho_{f} a^{\dagger}_{i}a_{i}).  \notag
\end{eqnarray}

The parameters $\rho_{ii,A}^{(0)}$ with $i=g,\,e$ and $\rho_{gd,A}^{(0)}$ can be
given as
\begin{eqnarray}
\rho_{gg,A}^{(0)}&=&\frac{(\gamma_{31}+\gamma_{32})\gamma_{21}(\Delta^2_{2}+%
\frac{1}{4}\Gamma^2_{2})+\Gamma_{2}\gamma_{21}\Omega^2 }{(\gamma_{31}+%
\gamma_{32})\gamma_{21}(\Delta^2_{2}+\frac{1}{4}\Gamma^2_{2})+\Omega^2%
\Gamma_{2}(2\gamma_{21}+\gamma_{32})},  \notag \\
\rho_{ee,A}^{(0)}&=&\frac{\gamma_{32}\Omega^2\Gamma_{2} }{%
(\gamma_{31}+\gamma_{32})\gamma_{21}(\Delta^2_{2}+\frac{1}{4}%
\Gamma^2_{2})+\Omega^2\Gamma_{2}(2\gamma_{21}+\gamma_{32})},  \notag \\
\rho_{gd,A}^{(0)}&=&i \frac{(\gamma_{31}+\gamma_{32})\gamma_{21}(i\Delta_{2}+%
\frac{1}{2}\Gamma_{2})\Omega^2\exp(i\phi)} {(\gamma_{31}+\gamma_{32})%
\gamma_{21}(\Delta^2_{2}+\frac{1}{4}\Gamma^2_{2})+\Omega^2\Gamma_{2}(2%
\gamma_{21}+\gamma_{32})},  \notag
\end{eqnarray}
where the decay rates are
\begin{eqnarray}
\Gamma_{1}&=&\gamma_{21}+\gamma_{22},  \notag \\
\Gamma_{2}&=&\gamma_{33}+\gamma_{32}+\gamma_{31},  \notag \\
\Gamma_{3}&=&\gamma_{33}+\gamma_{22}+\gamma_{21}+\gamma_{32}+\gamma_{31}.
\notag
\end{eqnarray}
The parameters are $\rho^{(0)}_{dd,A}=1-\rho^{(0)}_{ee,A}-\rho^{(0)}_{gg,A}$ and
$\rho_{dg,A}^{(0)}=[\rho_{gd,A}^{(0)}]^{*}$. Also
$\alpha_{1}$, $\beta_{i}$, $\mu_{i}$ and $\nu_{i}$ with $i=1,2$
are given by
\begin{eqnarray}
\alpha_{1}&=&\frac{g_{2}^{2}}{D_{1}}\left(i\Delta_{3}+\frac{\Gamma_{3}}{2}%
+i\omega_{2}\right), \\
\alpha_{2}&=&\frac{g_{1}^{2}}{D_{2}}\left(-i\omega_{21}+\frac{\Gamma_{1}}{2}%
-i\Delta_{1}\right), \\
\beta_{1}&=&\frac{g_{1}g_{2}}{D_{2}}\left(i\Delta_{3}+\frac{\Gamma_{3}}{2}%
-i\Delta_{1}\right), \\
\beta_{2}&=&\frac{g_{1}g_{2}}{D_{1}^{*}}\left(-i\omega_{2}+i\omega_{21}+%
\frac{\Gamma_{1}}{2}\right),
\end{eqnarray}
and
\begin{eqnarray}
\mu_{1}&=&i\Omega\frac{g_{2}^2}{D_{1}^{*}}\exp(-i\phi), \\
\mu_{2}&=&i\Omega\frac{g_{1}^2}{D_{2}^{*}}\exp(i\phi), \\
\nu_{1}&=&i\Omega\frac{g_{1}g_{2}}{D_{1}^{*}}\exp(i\phi), \\
\nu_{2}&=&i\Omega\frac{g_{1}g_{2}}{D_{2}^{*}}\exp(-i\phi).
\end{eqnarray}
The denominators $D_{1}$ and $D_{2}$ in the above equations are
\begin{eqnarray}
D_{1}&=&\left(i\omega_{2}-i\omega_{21}+\frac{\Gamma_{1}}{2}
\right)\left(i\omega_{2}+i\Delta_{3}+\frac{\Gamma_{3}}{2} \right)+\Omega^2,
\notag \\
D_{2}&=&\left(-i\omega_{21}-i\Delta_{1}+\frac{\Gamma_{1}}{2}
\right)\left(i\Delta_{3}-i\Delta_{1}+\frac{\Gamma_{3}}{2} \right)+\Omega^2,
\notag
\end{eqnarray}
with the detuning $\Delta_{3}=\omega_{dg}-\omega_{p}-\omega_{eg}$. Using the
master equation for the reduced density matrix of the two cavity modes, we
have
\begin{eqnarray}
\frac{\partial \langle a_{1}\rangle}{\partial t}&=&-i\Delta_{1}\langle
a_{1}\rangle-\kappa_{1}\langle a_{1}\rangle +\frac{g_{1}^2}{D_{2}}%
\alpha_{2}[\rho_{dd,A}^{(0)}-\rho_{ee,A}^{(0)}]\langle a_{1}\rangle  \notag
\\
&-&\mu_{2}^{*}\rho_{gd,A}^{(0)}\langle
a_{1}\rangle-\nu_{1}^{*}[\rho_{ee,A}^{(0)}-\rho_{gg,A}^{(0)}] \langle
a_{2}^{\dagger}\rangle  \notag \\
&-&\frac{g_{1}g_{2}}{D_{1}}\beta_{2}\rho_{dg,A}^{(0)}\langle
a_{2}^{\dagger}\rangle,  \label{eq:11} \\
\frac{\partial \langle a_{2}\rangle}{\partial t}&=&-i\omega_{2}\langle
a_{2}\rangle-\kappa_{2}\langle a_{2}\rangle +\frac{g_{2}^2}{D_{1}^{*}}%
\alpha_{1}^{*}[\rho_{ee,A}^{(0)}-\rho_{gg,A}^{(0)}]\langle a_{2}\rangle
\notag \\
&-&\mu_{1}\rho_{gd,A}^{(0)}\langle a_{2}\rangle-\frac{g_{1}g_{2}}{D_{2}^{*}}%
\beta_{1}^{*} \rho_{dg,A}^{(0)}\langle a_{1}^{\dagger}\rangle  \notag \\
&+&\nu_{2}[\rho_{ee,A}^{(0)}-\rho_{dd,A}^{(0)}]\langle
a^{\dagger}_{1}\rangle.  \label{eq:12}
\end{eqnarray}
It is clear that the equations of motion for the expectation values $\langle
a^{\dagger}_{1}\rangle$ and $\langle a^{\dagger}_{2}\rangle$ can be obtained
by taking the conjugates of Eqs.~(\ref{eq:11}) and (\ref{eq:12}), respectively. We then can also
obtain
\begin{eqnarray}
\frac{\partial \langle a^{\dagger}_{1}a_{1}\rangle}{\partial t}%
&=&-2\kappa_{1}\langle
a^{\dagger}_{1}a_{1}\rangle+\left\{\mu_{2}^{*}\rho_{gd,A}^{(0)}+\alpha_{2}%
\rho_{dd,A}^{(0)}+\text{H.c.}\right\} \\
&+&\left\{\left(\mu_{2}^{*}\rho_{gd,A}^{(0)}+\alpha_{2}\rho_{dd,A}^{(0)}-%
\alpha_{2}\rho_{ee,A}^{(0)}\right)\langle a^{\dagger}_{1}a_{1}\rangle+\text{%
H.c.}\right\}  \notag \\
&+&\left\{\left[\nu_{1}\rho_{gg,A}^{(0)}-\nu_{1}\rho_{ee,A}^{(0)}+\beta_{2}%
\rho_{gd,A}^{(0)}\right]\langle a_{1}a_{2}\rangle+\text{H.c.}\right\}, \label{eq:13}
\notag \\
\frac{\partial \langle a^{\dagger}_{2}a_{2}\rangle}{\partial t}%
&=&-2\kappa_{2}\langle
a^{\dagger}_{2}a_{2}\rangle+(\alpha_{1}+\alpha_{1}^{*})\rho_{ee,A}^{(0)} \\
&+&\left\{\left[\alpha_{1}\rho_{ee}^{(0)}-\alpha_{1}\rho_{gg}^{(0)}-\mu_{1}%
\rho_{gd,A}^{(0)}\right]\langle a^{\dagger}_{2}a_{2}\rangle+\text{H.c.}%
\right\}  \notag \\
&-&\left\{\left[\beta_{1}\rho_{gd,A}^{(0)}+\nu_{2}^{*}\rho_{dd,A}^{(0)}-%
\nu_{2}^{*}\rho_{ee,A}^{(0)}\right]\langle a_{1}a_{2}\rangle+\text{H.c.}%
\right\}, \label{eq:14} \notag \\
\frac{\partial \langle a^{\dagger}_{1}a^{\dagger}_{2}\rangle}{\partial t}%
&=&i(\omega_{1}+\omega_{2}-\omega_{d})\langle
a^{\dagger}_{1}a^{\dagger}_{2}\rangle-(\kappa_{1}+\kappa_{2})\langle
a^{\dagger}_{1}a^{\dagger}_{2}\rangle \\
&+&\left\{\alpha_{1}\left[\rho_{ee,A}^{(0)}-\rho_{gg,A}^{(0)}\right]%
+\left(\mu_{2}-\mu_{1}^{*}\right)\rho_{dg,A}^{(0)}\right\}\langle
a^{\dagger}_{1}a^{\dagger}_{2}\rangle  \notag \\
&+&\alpha_{2}\left[\rho_{dd,A}^{(0)}-\rho_{ee,A}^{(0)}\right]\langle
a^{\dagger}_{1}a^{\dagger}_{2}\rangle-\beta_{1}\rho_{gd,A}^{(0)}-\nu_{2}^{*}%
\rho_{dd,A}^{(0)}  \notag \\
&-&\left[\beta_{1}\rho_{13,A}+\nu_{2}^{*}\rho_{dd,A}^{(0)}-\nu_{2}^{*}%
\rho_{ee,A}^{(0)}\right]\langle a^{\dagger}_{1}a_{1}\rangle  \notag \\
&+&\left\{\nu_{1}\left[\rho_{gg,A}^{(0)}-\rho_{ee,A}^{(0)}\right]%
+\beta_{2}\rho_{gd,A}^{(0)}\right\}\langle
a^{\dagger}_{2}a_{2}\rangle-\nu_{1}\rho_{ee,A}^{(0)}. \label{eq:15}  \notag
\end{eqnarray}
The steady-state average photon number can be
approximately obtained from the above equations. At the working point, we expect $\Delta_2=2\pi\times30\,$MHz,
$\Omega=2\pi\times900\,$MHz, $\phi=0$, $\omega_{eg}=2\pi\times11.4979\,$GHz and $\omega_{de}=2\pi\times6.5376\,$GHz. Using the dissipation terms of the two modes and the artificial atom (the dissipation terms of the atom are extracted below), we obtain $\langle a_{1}^+a_1\rangle\approx5$ and $\langle a_{2}^+a_2\rangle\approx2$. Furthermore, both $\langle a_{1}^+a_1\rangle$ and $\langle a_{2}^+a_2\rangle$ decrease when we change the pumping power or pumping frequency away from these optimal parameters, in good agreement with the experiment.

\subsection{Quantum correlation between two lasing modes}

We use a criterion derived in \cite{Duan2000} to study the quantum correlation between
the two lasing modes in two-mode correlated emission lasing. According to this criterion, a necessary and sufficient condition for a system to be in a quantum entanglement state is for the sum of the quantum fluctuations of the two Einstein-Podolsky-Rosen (EPR)-like operators $\hat{u}$ and $\hat{v}$ of the two modes to satisfy the inequality
\begin{equation}\label{entanglecondition}
(\Delta\hat{u})^2+(\Delta\hat{v})^2<2.
\end{equation}
Here, $\hat{u}=\hat{x_1}+\hat{x_2}$, $\hat{v}=\hat{p_1}-\hat{p_2}$, $x_j=(a_j+a^+_j)/\sqrt{2}$ and $p_j=(a_j-a^+_j)/\sqrt{2}i$ (with $j=1,2$) are the quadratures for the two lasing modes.

After substituting the definitions of $\hat{u}$ and $\hat{v}$ in Eq.~(\ref{entanglecondition}), we obtain
\begin{eqnarray}
 (\Delta\hat{u})^2+(\Delta\hat{v})^2 &=& 2[1+\langle a_1a_1^+\rangle+\langle a_2a_2^+
 \rangle+\langle a_1a_2^+\rangle+\langle a_1^+a_2^+\rangle \notag \\
                                     & &-\langle a_1\rangle\langle a_1^+\rangle                                     -\langle a_2\rangle\langle a_2^+\rangle-\langle a_1\rangle\langle a_2\rangle-\langle a_1^+\rangle\langle a_2^+\rangle].
 \end{eqnarray}

\begin{figure}
\center
\includegraphics[scale=0.7]{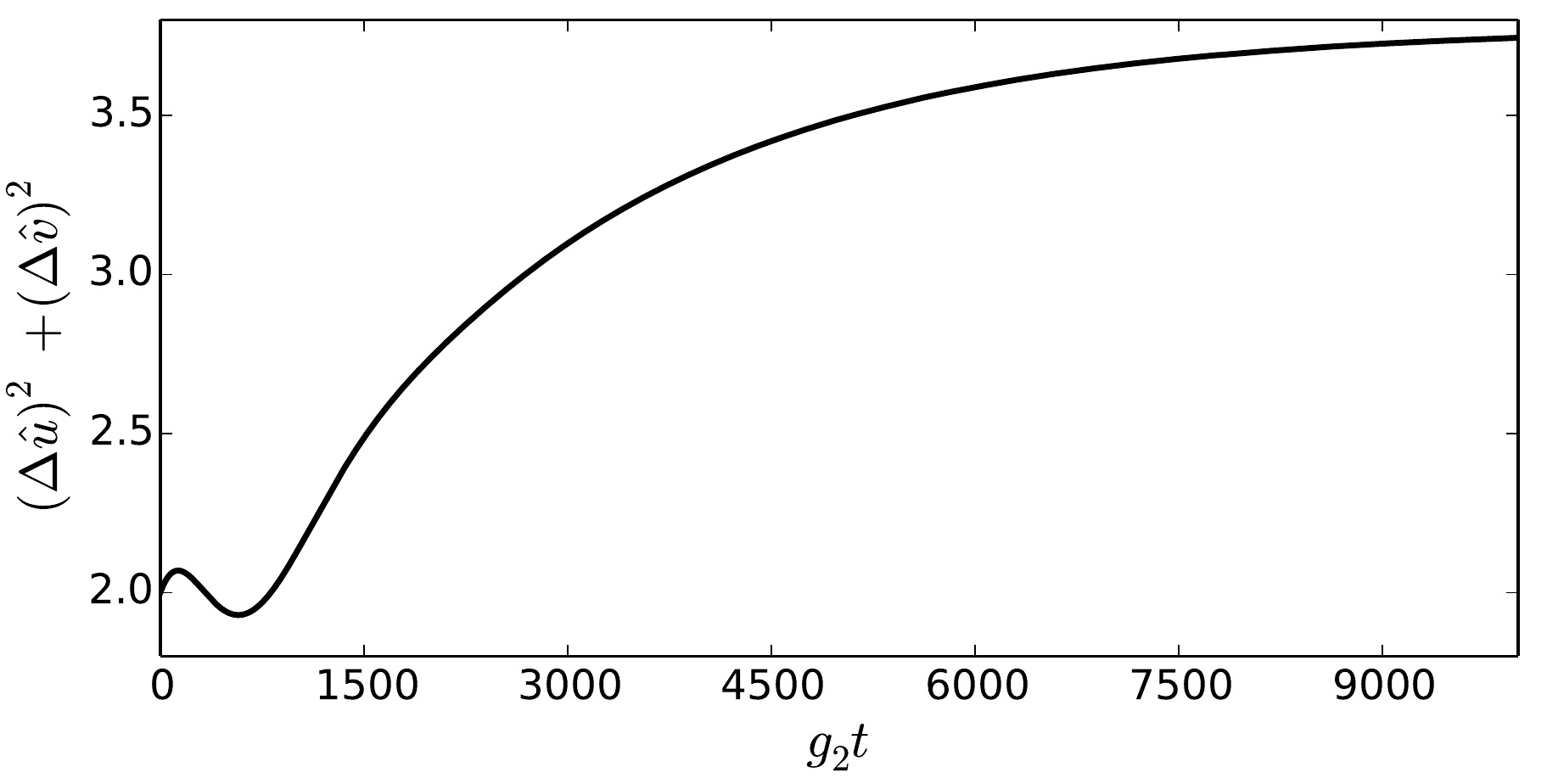}
\caption{Time dependence of $(\Delta\hat{u})^2+(\Delta\hat{v})^2$ for initial vacuum states in terms of the normalized time $g_2t$.}\label{quantumCorrelation}
\end{figure}

We can calculate $(\Delta\hat{u})^2+(\Delta\hat{v})^2$ versus time using Eqs.~(\ref{eq:11})-(\ref{eq:15}), their conjugate equations and the experimental parameters. The data are presented in Fig.~\ref{quantumCorrelation}. We thus show the possible existence of quantum correlation between the two lasing fields in our CEL system.
\subsection{Phase diffusions}

To study the phase noise, we can convert Eq.~(\ref{eq:7}) into an equivalent
Fokker-Planck equation for the P-representation $P\equiv P(q_{1},q_{1}^{*},
q_{2},q_{2}^{*})$ using the following relations:
\begin{eqnarray}
a_{i}\rho_{f}&\leftrightarrow& q_{i}P , \\
\rho_{f}a^{\dagger}_{i}&\leftrightarrow& q^{*}_{i}P , \\
a^{\dagger}_{i}\rho_{f}&\leftrightarrow& \left(q^{*}_{i}-\frac{\partial }{%
\partial q_{i}}\right)P , \\
\rho_{f}a_{i}&\leftrightarrow& \left(q_{i}-\frac{\partial}{\partial q^{*}_{i}%
}\right)P , \\
a^{\dagger}_{i}a_{i}\rho_{f}&\leftrightarrow& \left(q^{*}_{i}-\frac{\partial
}{\partial q_{i}}\right)q_{i}P, \\
\rho_{f}a^{\dagger}_{i}a_{i}&\leftrightarrow& \left(q_{i}-\frac{\partial}{%
\partial q^{*}_{i}}\right)q^{*}_{i}P, \\
a_{i}a^{\dagger}_{i}\rho_{f}&\leftrightarrow& q_{i}\left(q^*_{i}-\frac{%
\partial}{\partial q_{i}}\right)P, \\
\rho_{f}a_{i}a^{\dagger}_{i}&\leftrightarrow& q^{*}_{i}\left(q_{i}-\frac{%
\partial}{\partial q^{*}_{i}}\right)P, \\
a^{\dagger}_{i}\rho_{f}a_{i}&\leftrightarrow& \left(q^{*}_{i}-\frac{\partial%
}{\partial q_{i}}\right)\left(q_{i}-\frac{\partial}{\partial q^{*}_{i}}%
\right)P.
\end{eqnarray}
Thus, the Fokker-Planck equation is
\begin{eqnarray}
\frac{\partial P}{\partial t}&=&\left[\left(\alpha_{2}\rho_{ee,A}^{(0)}+%
\kappa_{1}+i\Delta_{1}\right)\frac{\partial}{\partial q_{1}}q_{1}+\text{H.c.}%
\right]P  \notag \\
&+&\left[\left(\alpha^*_{1}\rho_{gg,A}^{(0)}+\kappa_{2}+\mu_{1}%
\rho_{gd,A}^{(0)}-i\omega_{2}\right)\frac{\partial}{\partial q_{2}}q_{2}+%
\text{H.c.}\right]P  \notag \\
&-&\left[\left(\beta_{2}^{*}\rho_{dg,A}^{(0)}+\nu_{1}^{*}\rho_{gg,A}^{(0)}%
\right)\frac{\partial}{\partial q_{1}}q_{2}^{*}+\nu_{2}\rho_{ee,A}^{(0)}%
\frac{\partial}{\partial q_{2}}q_{1}^{*}+\text{H.c.}\right]P  \notag \\
&-&\left[\alpha_{1}^*\rho_{ee,A}^{(0)}\left(\frac{\partial}{\partial q_{2}}%
q_{2}-\frac{\partial^2}{\partial q_{2}\partial q_{2}^{*}}\right)+\text{H.c.}%
\right]P  \notag \\
&-&\left[\left(\alpha_{2}\rho_{dd,A}^{(0)}+\mu^{*}_{2}\rho_{gd,A}^{(0)}%
\right)\left(\frac{\partial}{\partial q_{1}}q_{1}-\frac{\partial^2}{\partial
q_{1}\partial q_{1}^{*}}\right)+\text{H.c.}\right]P  \notag \\
&+&\left[\left(\beta^{*}_{1}\rho_{dg,A}^{(0)}+\nu_{2}\rho_{dd,A}^{(0)}%
\right)\left(\frac{\partial}{\partial q_{2}}q^*_{1}-\frac{\partial^2}{%
\partial q_{1}\partial q_{2}}\right)+\text{H.c.}\right]P  \notag \\
&+&\left[\nu^*_{1}\rho_{ee,A}^{(0)}\left(\frac{\partial}{\partial q_{1}}%
q^*_{2}-\frac{\partial^2}{\partial q_{1}\partial q_{2}}\right)+\text{H.c.}%
\right]P.
\end{eqnarray}

We now define the polar
coordinates $r_{i}\ $and\ $\theta _{i}$ $\left( i=1,2\right) $ via relations as in Ref.~\cite{Scully-book}
\begin{equation*}
q_{i}=r_{i}\exp \left( i\theta _{i}\right)
\end{equation*}%
We define the difference and mean angle variables as%
\begin{eqnarray*}
\theta  &=&\theta _{1}-\theta _{2}, \\
\eta  &=&\frac{\theta _{1}+\theta _{2}}{2},
\end{eqnarray*}%
and then we have%
\begin{equation*}
\frac{\partial }{\partial q_{l}}=\frac{\exp \left( -i\theta _{l}\right) }{2}%
\left[ \frac{\partial }{\partial r_{l}}+\frac{1}{ir_{l}}\frac{\partial }{%
\partial \theta _{l}}\right],
\end{equation*}%
\begin{eqnarray*}
\frac{\partial }{\partial \theta _{1}} &=&\frac{1}{2}\frac{\partial }{%
\partial \eta }+\frac{\partial }{\partial \theta } \\
\frac{\partial }{\partial \theta _{2}} &=&\frac{1}{2}\frac{\partial }{%
\partial \eta }-\frac{\partial }{\partial \theta }.
\end{eqnarray*}%
Under the lasing conditions, the variation in the photon number is
small and we can consider that $\frac{\partial }{\partial r_{i}}=0$. Then,
we have%
\begin{eqnarray*}
\frac{\partial }{\partial q_{1}} &\approx &\frac{\exp \left( -i\theta
_{1}\right) }{2}\frac{1}{ir_{1}}\frac{\partial }{\partial \theta _{1}} \\
\frac{\partial }{\partial q_{2}} &\approx &\frac{\exp \left( -i\theta
_{2}\right) }{2}\frac{1}{ir_{2}}\frac{\partial }{\partial \theta _{2}}
\end{eqnarray*}%
and%
\begin{equation*}
\frac{dP}{dt}=\frac{\partial }{\partial \theta }D_{\theta }P+\frac{\partial
}{\partial \eta }D_{\eta }P+\frac{\partial ^{2}}{\partial \theta ^{2}}%
D_{\theta \theta }P+\frac{\partial ^{2}}{\partial \eta ^{2}}D_{\eta \eta }P+%
\frac{\partial ^{2}}{\partial \theta \partial \eta }D_{\theta \eta }P
\end{equation*}%
with%
\begin{eqnarray*}
&&D_{\theta \theta }=\left[ \frac{1}{4r_{2}^{2}}\alpha _{1}\rho
_{ee,A}^{\left( 0\right) }+\frac{1}{4r_{1}^{2}}\left( \alpha _{2}\rho
_{dd,A}^{\left( 0\right) }+\mu _{2}\rho _{dg,A}^{\left( 0\right) }\right) +%
\text{H.c.}\right]  \\
&&-\left\{ \frac{1}{4r_{2}r_{1}}\left[ \beta _{1}\rho _{gd,A}^{\left(
0\right) }+\nu _{2}^{\ast }\rho _{dd,A}^{\left( 0\right) }+\nu _{1}\rho
_{ee,A}^{\left( 0\right) }\right] \exp \left[ i2\eta \right] +\text{H.c.}%
\right\}
\end{eqnarray*}%
\begin{eqnarray*}
D_{\eta \eta } &=&\left[ \frac{1}{16r_{2}^{2}}\alpha _{1}\rho
_{ee,A}^{\left( 0\right) }+\frac{1}{16r_{1}^{2}}\left( \alpha _{2}\rho
_{dd,A}^{\left( 0\right) }+\mu _{2}\rho _{dg,A}^{\left( 0\right) }\right) +%
\text{H.c.}\right]  \\
&&+\left\{ \frac{1}{16r_{2}r_{1}}\left[ \beta _{1}\rho _{gd,A}^{\left(
0\right) }+\nu _{2}^{\ast }\rho _{dd,A}^{\left( 0\right) }+\nu _{1}\rho
_{ee,A}^{\left( 0\right) }\right] \exp \left[ i2\eta \right] +\text{H.c.}%
\right\}
\end{eqnarray*}

\begin{equation*}
D_{\theta \eta }=\frac{1}{4r_{1}^{2}}\left[ \alpha _{2}\rho _{33,A}^{\left(
0\right) }+\mu _{2}^{\ast }\rho _{13,A}^{\left( 0\right) }\right] -\frac{1}{%
4r_{2}^{2}}\alpha _{1}\rho _{22,A}^{\left( 0\right) }+\text{H.c.}
\end{equation*}%
\begin{eqnarray*}
D_{\eta } &=&\frac{1}{2}\left( \Delta _{1}+\omega _{2}\right)  \\
&&+\frac{i}{4}\left[ \left( \alpha _{2}^{\ast }+\alpha _{1}^{\ast }\right)
\rho _{22,A}^{\left( 0\right) }+\left( \alpha _{1}\rho _{11,A}^{\left(
0\right) }-\mu _{1}\rho _{13,A}^{\left( 0\right) }\right) -\text{H.c.}\right]
\\
&&+\left[ \frac{i}{4}\left( \alpha _{2}\rho _{33,A}^{\left( 0\right) }-\mu
_{2}\rho _{31,A}^{\left( 0\right) }\right) +\text{H.c.}\right]  \\
&&+\left[ i\frac{r_{2}}{4r_{1}}\left[ \nu _{1}\rho _{22,A}^{\left( 0\right)
}-\beta _{2}\rho _{13,A}^{\left( 0\right) }-\nu _{1}\rho _{11,A}^{\left(
0\right) }\right] \exp \left[ i2\eta \right] +\text{H.c.}\right]  \\
&&+\left[ i\frac{r_{1}}{4r_{2}}\left[ \beta _{1}\rho _{13,A}^{\left(
0\right) }+\nu _{2}^{\ast }\rho _{33,A}^{\left( 0\right) }-\nu _{2}^{\ast
}\rho _{22,A}^{\left( 0\right) }\right] \exp \left[ i2\eta \right] +\text{%
H.c.}\right]  \\
&&+\left[ i\frac{1}{4r_{2}r_{1}}\left[ \beta _{1}^{\ast }\rho
_{31,A}^{\left( 0\right) }+\nu _{2}\rho _{33,A}^{\left( 0\right) }+\nu
_{1}^{\ast }\rho _{22,A}^{\left( 0\right) }\right] \exp \left[ -i2\eta %
\right] +\text{H.c.}\right]
\end{eqnarray*}%
\begin{eqnarray*}
D_{\theta } &=&\Delta _{1}+\omega _{2} \\
&&+\left\{ \frac{i}{2}\left[ \alpha _{2}\left( \rho _{33,A}^{\left( 0\right)
}-\rho _{22,A}^{\left( 0\right) }\right) -\mu _{2}\rho _{31,A}^{\left(
0\right) }\right] +\text{H.c.}\right\}  \\
&&+\left\{ \frac{i}{2}\left[ \mu _{1}\rho _{13,A}^{\left( 0\right) }-\alpha
_{1}\left( \rho _{11,A}^{\left( 0\right) }-\rho _{22,A}^{\left( 0\right)
}\right) \right] +\text{H.c.}\right\}  \\
&&+\left\{ i\frac{r_{1}}{2r_{2}}\left[ \nu _{2}^{\ast }\left( \rho
_{22,A}^{\left( 0\right) }-\rho _{33,A}^{\left( 0\right) }\right) -\beta
_{1}\rho _{13,A}^{\left( 0\right) }\right] \exp \left( i2\eta \right) +\text{%
H.c.}\right\}  \\
&&+\left\{ i\frac{r_{2}}{2r_{1}}\left[ \nu _{1}\left( \rho _{22,A}^{\left(
0\right) }-\rho _{11,A}^{\left( 0\right) }\right) -\beta _{2}\rho
_{13,A}^{\left( 0\right) }\right] \exp \left[ i2\eta \right] +\text{H.c.}%
\right\}
\end{eqnarray*}
Here, $D_{\theta\theta}$ and $D_{\eta\eta}$ characterize the phase diffusions of $\theta$ and $\eta$, respectively.

From the Fokker-Planck equation, we can derive the equations of motion for the phase difference $\theta$ and phase sum $\eta$ as follows:
\begin{eqnarray}
\frac{d \theta}{d t}&=&D_{\theta},\\
\frac{d \eta }{d t}&=& D_{\eta}.
\end{eqnarray}
Here, $D_{\theta}=0$ or $D_{\eta}=0$ means that the phase is locked. However, $D_{\theta\theta}\leq0$ or $D_{\eta\eta}\leq0$ means that the phase diffusions are suppressed. We substitute all the dissipation terms, $r_{1}=\sqrt{\langle a_{1}^+a_1\rangle}=\sqrt{5}$ and $r_{2}=\sqrt{\langle a_{2}^+a_2\rangle}=\sqrt{2}$ to calculate $D_{\eta\eta}$. Finally, we find that $D_{\eta\eta}<0$, which means that the mutual phase-diffusion noise is suppressed.
\section{Methods}
\subsection{Artificial atom}

As shown in Fig.~1(A), (B) and (C) of the main text, the single three-level artificial atom with cyclic transitions~\cite{Liu2005}  is based on a \textquotedblleft tunable gap flux qubit\textquotedblright   circuit~\cite{Paauw2009,Zhu2010} capacitively coupled to a multimode transmission line resonator (TLR) via $C_c$ \cite{Inomata2012}.
The effective Josephson and charging energies of each junction shown in blue in Fig. 1C are $E_{j}=h\times88\,$GHz and $E_{c}=h\times3.1\,$GHz (where $E_c = e^2/2C$ for the junction capacitance $C$), $\alpha=1.02$ (ratio of the total junction capacitance marked in red in Fig. 1C to $C$), $C_{c}=3.1\,$fF, and the area ratio of the main loop to the $\alpha$-loop is 6.1.
The niobium (Nb) TLR is patterned by the dry etching of a 50-nm-thick Nb thin film deposited on a thermally oxidized silicon substrate. The Nb TLR structure has a center conductor width of  $10\,\mu\rm m$ and gaps of $6\,\mu\rm m$, resulting in a wave impedance of about $50\,\Omega$ with a total length of the central conductor of $L\simeq10.2\,\rm mm$. Our experiment is carried out in a dilution refrigerator at a base temperature of about $30\,$mK.

The artificial atom is fabricated near the voltage antinode of the resonator by the electron-beam lithography and Al/AlOx/Al shadow evaporation techniques.
The energies of the three lowest levels of the atom (ground $|g\rangle$, first excited state $|e\rangle$ and second excited state $|d\rangle$) depend on the external magnetic flux, reaching minima for the transitions from $|g\rangle$ to $|e\rangle$ and $|d\rangle$ at half-integer flux quanta $\Phi_N = (N+\frac{1}{2})\Phi_{0}$ (where $N$ is an integer).
Compared with superconducting systems based on the conventional flux qubit geometry \cite{Mooij1999,Wal2000}, our atom has additional tunability due to the implementation of an $\alpha$-loop -- a dc SQUID, which allows the qubit energy gap $\Delta$ to be tuned by controlling the magnetic flux through the SQUID loop. When the biased flux $\Phi$ is close to $\Phi_N$, the two lowest energy eigenstates ($|g\rangle$ and $|e\rangle$) of the artificial atom are in the superposition of the clockwise and anticlockwise circulating current states of the main loop. The atomic transition frequency in  the vicinity of $\Phi_N$ in the first order is described by $\hbar\omega_{eg}\approx\sqrt{(2I_{p}\delta\Phi)^{2}+\Delta(\Phi_N)^{2}}$, where $\delta\Phi = \Phi-\Phi_N$ and $I_{p}$ is the persistent current in the main loop. Here, we neglect the weak dependence of $\Delta$ on $\Phi$ when $\delta\Phi \ll \Phi_0$ by assuming that $\Delta(\Phi)\approx\Delta(\Phi_N)$.
By choosing $\Phi_N$ and $\delta\Phi$, we can adjust both frequencies, $\omega_{eg}$ and $\omega_{de}$, and bring them close to the resonance of the two lowest modes of the resonator, which are two independent fixed-frequency oscillators with the frequencies $\omega_1$ and $\omega_2$.

\subsection{Extracting the dissipation terms of the artificial atom}
%The dashed lines in Fig.~\ref{Spectroscopywithoutfitting}D, Fig.~\ref{Spectroscopywithoutfitting}E and Fig.~\ref{Spectroscopywithoutfitting}F are the fitting data with the full coupled system Hamiltonian (Eq. (1) in Ref. \cite{Inomata2012}).

\begin{figure}
\includegraphics[scale=0.4]{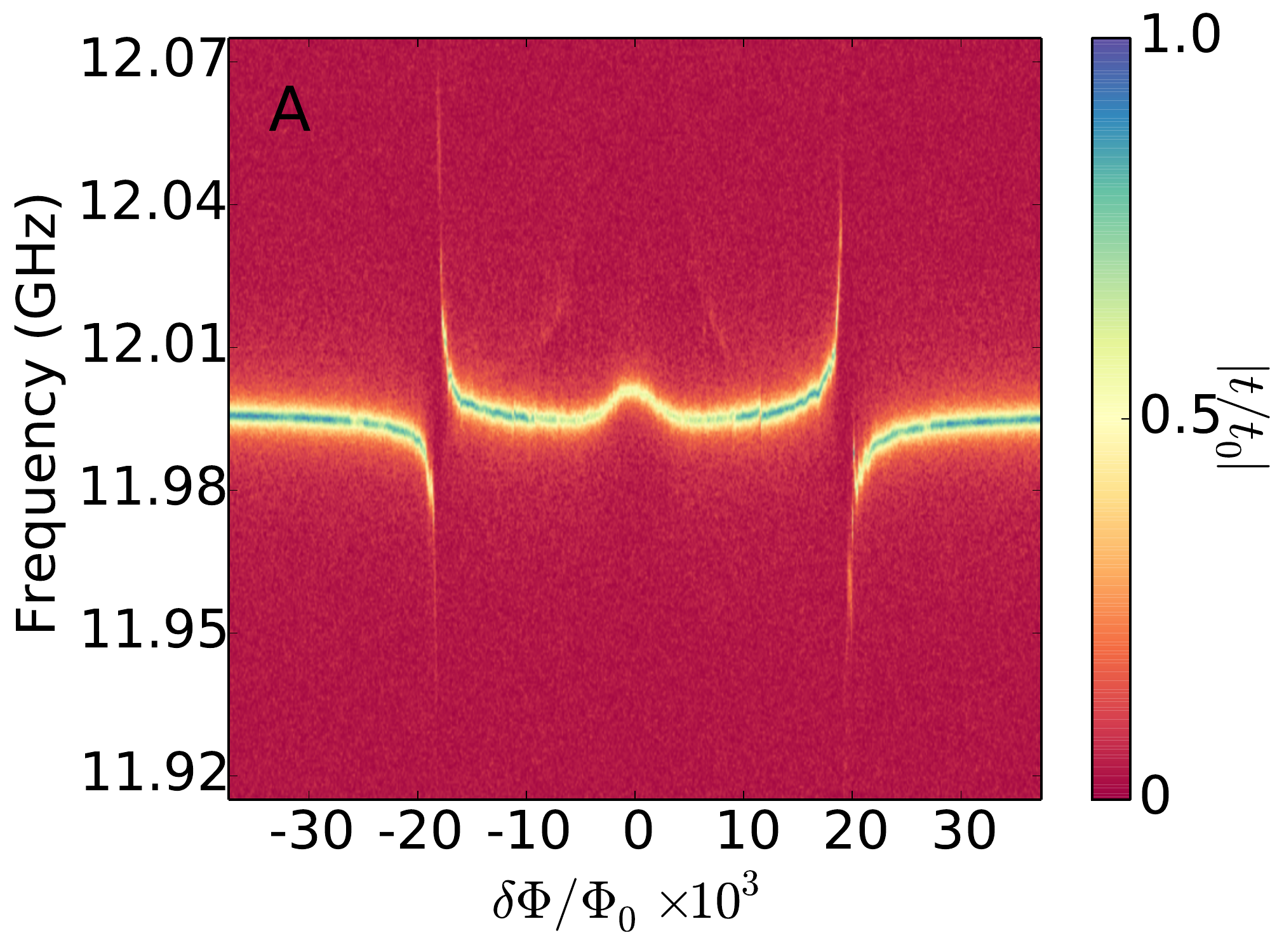}
\includegraphics[scale=0.4]{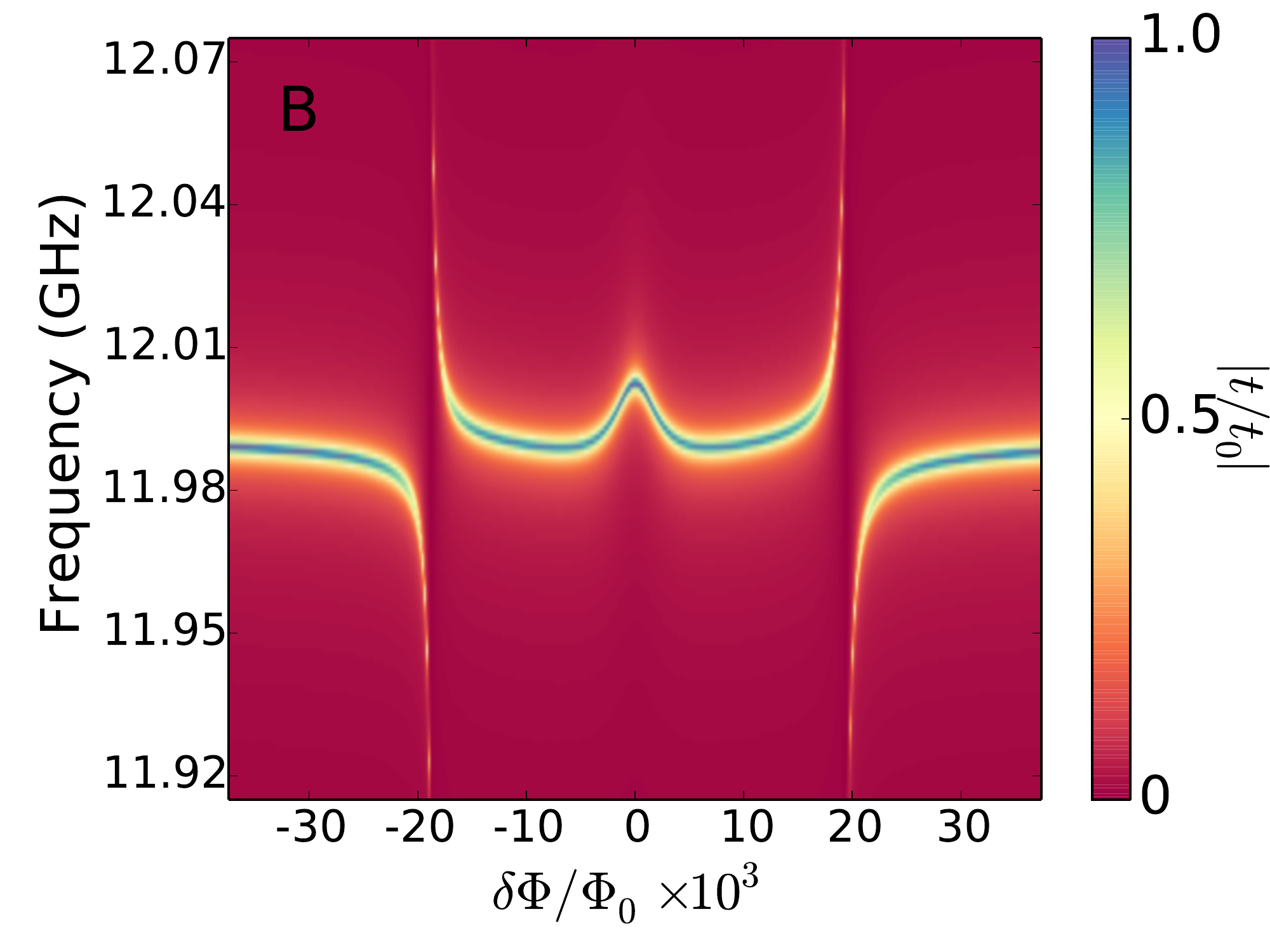}
\caption{Experimentally measured (A) and theoretical simulations (B) of the transmission spectrum for the second mode. From the simulations, we can extract the dissipation terms of the three-level artificial atom. }\label{Simulationoftransmission}
\end{figure}

As shown in Fig.~\ref{Simulationoftransmission}, we simulate the transmission spectrum for the second mode of the resonator around $12\,$GHz in Fig. 2B to extract the dissipation terms of the three-level artificial atom at the CEL working point.
For a probe field with the frequency $\omega_d$ driving the resonator $H_{d}=i\frac{\Omega_d}{2}(a_{2}^{+}e^{-i\omega_dt}-a_{2}e^{i\omega_dt})$, the total Hamiltonian in the rotating frame at the frequency $\omega_{d}$ is
\begin{eqnarray}\label{Rotatingframe}
  H&=&\hbar(\omega_{2}-\omega_{d})a_{2}^{+}a_{2}+\frac{\hbar}{2}(\omega_{dg}-\omega_d)\sigma_{dd}+\frac{\hbar}{2}(\omega_{eg}-\omega_d)\sigma_{ee}+i\frac{\Omega_d}{2}(a_{2}^{+}-a_{2}) \\ \nonumber
   & &+g_{4}(a_{2}^{+}\sigma_{de}+a_{2}\sigma_{ed})+\hbar g_{2}\left( a_{2}^{\dag }\sigma _{eg}+a_{2}\sigma_{ge}\right)+g_{5}(a_{2}^{+}\sigma_{dg}+a_{2}\sigma_{gd}),
\end{eqnarray}
where the amplitude is $\Omega_d=\hbar\kappa_{2}\sqrt{N}$ in terms of the photon number $N$ created by the driving field. To extract the dissipation terms of the artificial atom, we can simulate the transmission spectrum in Fig.~2B by solving the stationary master equation $\dot{\rho}=0$ with the Hamiltonian in Eq.~(\ref{Rotatingframe}) and the decay rate of the second mode. The transmission amplitude $t$ in the second-harmonic mode of the resonator can be written as
\begin{equation}\label{transmission}
 t=-\frac{i\kappa_{2}\langle a_{2} \rangle}{\Omega_d},
\end{equation}
where $\langle a_2 \rangle$ is the expectation value of the photon field in the second-harmonic mode of the resonator.
In the weak driving limit, Eq.~(\ref{transmission}) returns back to Eq.~(9) in Ref.~\cite{Oelsner2010}.
We obtain the parameters $\gamma_{21}\approx2\pi\times1.5\,$MHz, $\gamma_{22}=2\pi\times6\,$MHz, $\gamma_{32}=2\pi\times3\,$MHz, $\gamma_{31}=2\pi\times8\,$MHz and $\gamma_{33}=2\pi\times4\,$MHz.
We consider that the small deviation between the experiment and the theoretical simulation is due to the fact that Eq.~(\ref{Rotatingframe}) is not the full system Hamiltonian.
% ----------------------FIGURE1 -----------------------------------

\subsection{Interactions between the artificial atom and different resonator modes}

\begin{figure}
\center
\includegraphics[scale=0.5]{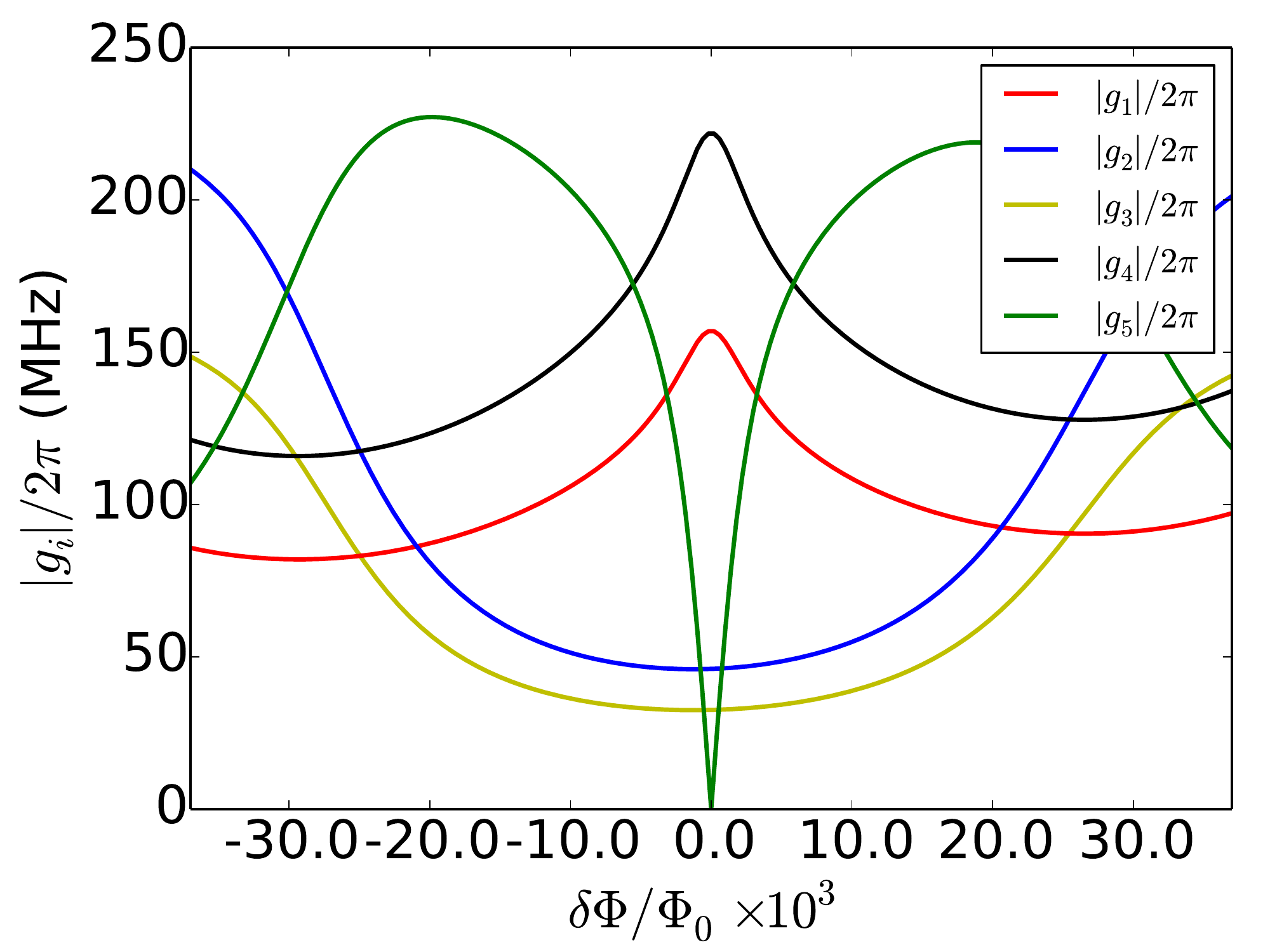}
\caption{Theoretical calculation of the coupling strength between different atomic transitions and cavity modes. $g_1$, $g_2$, $g_3$, $g_4$ and $g_5$ correspond to the interactions $|0 d 0\rangle <=> |1 e 0\rangle$, $|0 e 0\rangle <=> |0 g 1\rangle$, $|0 e 0\rangle <=> |1 g 0\rangle$, $|0 d 0\rangle <=> |0 e 1\rangle$ and the coupling strength between $|0 d 0\rangle <=> |0 g 0\rangle$ and the third mode of the resonator, respectively. }\label{Couplingstrength}
\end{figure}

Using the sample circuit parameters, we can calculate the coupling strength between the different transitions of the atom and the different modes of the resonator as shown in Fig.~\ref{Couplingstrength}. In the figure, $g_1$, $g_2$, $g_3$, $g_4$ and $g_5$ correspond to the interactions $|0 d 0\rangle <=> |1 e 0\rangle$, $|0 e 0\rangle <=> |0 g 1\rangle$, $|0 e 0\rangle <=> |1 g 0\rangle$, $|0 d 0\rangle <=> |0 e 1\rangle$ and the coupling strength between $|0 d 0\rangle <=> |0 g 0\rangle$ and the third mode of the resonator, respectively. At the working point, $g_{1}=2\pi\times90\,$MHz, $g_{2}=2\pi\times78\,$MHz and $g_{5}=2\pi\times225\,$MHz.
% ----------------------FIGURE1 -----------------------------------
%% ----------------------FIGURE -----------------------------------

\subsection{Experimental setups for the transmission and emission measurements}

\begin{figure}
\center
\includegraphics[scale=0.6]{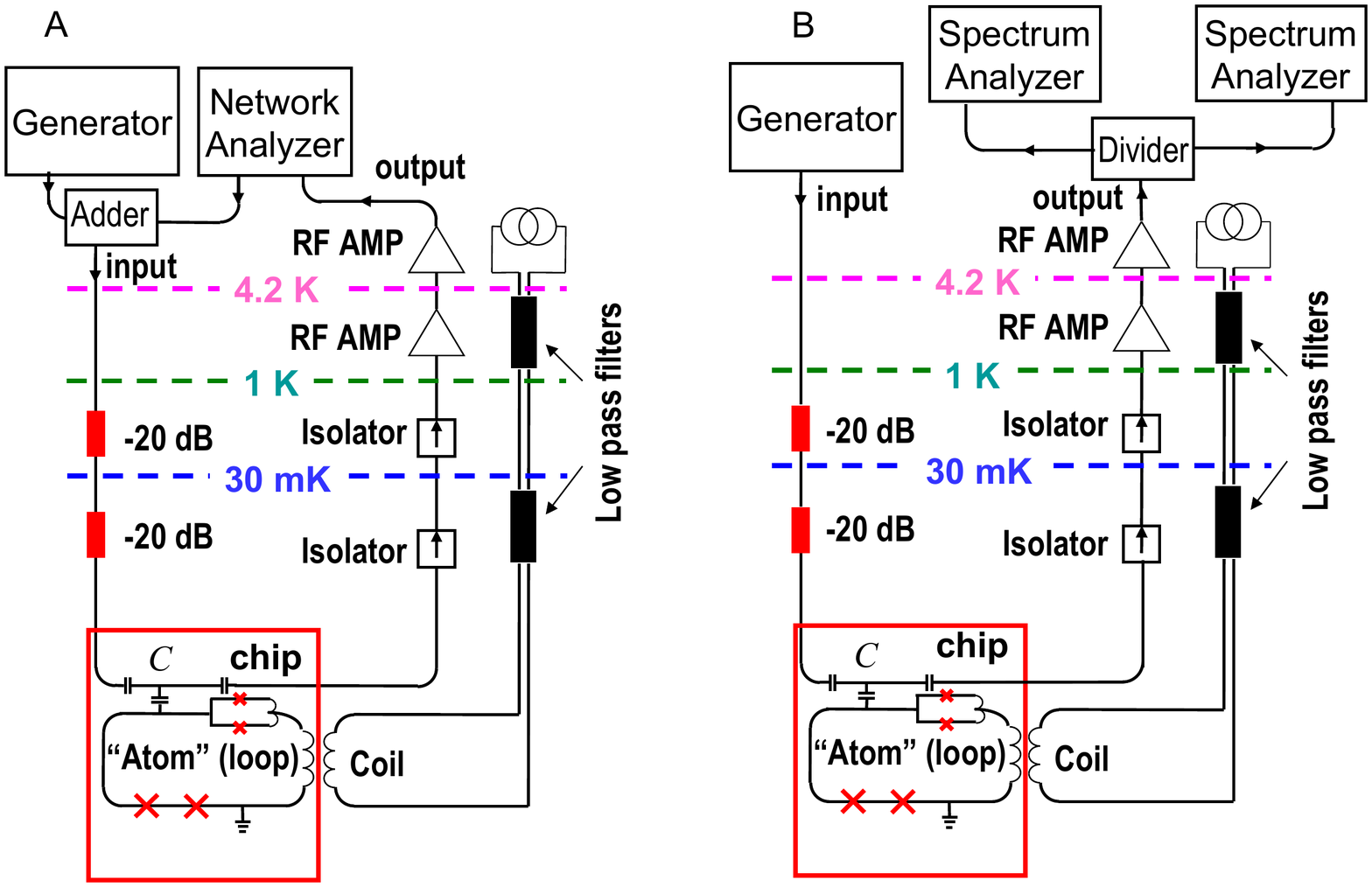}
\caption{({\bf A}) The experimental setup for measuring transmission spectrum and energy spectroscopy of the coupled system. ({\bf B}) The experimental setup for measuring two-mode CEL emission spectrum.}\label{TransmissionandEmissionSetup}
\end{figure}

The experimental setups used to measure the transmission and emission are presented in Fig.~\ref{TransmissionandEmissionSetup} A and Fig.~\ref{TransmissionandEmissionSetup}B, respectively.
% ----------------------FIGURE -----------------------------------
%% ----------------------FIGURE -----------------------------------

\begin{figure}
\center
\includegraphics[scale=0.8]{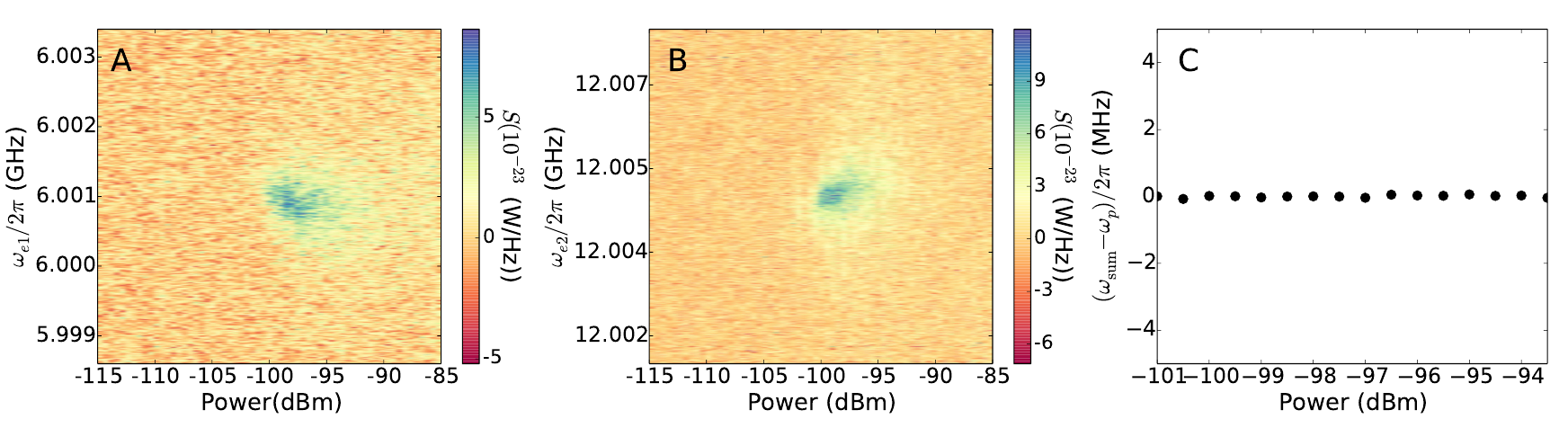}
\caption{Sum of the two center frequencies of emission spectra is equal to the pumping frequency with changing the pumping power. Emission spectra from the first mode ({\bf A}) and second mode ({\bf B}) with changing the pumping power. ({\bf C}) Sum of the two center frequencies of the emission spectra in ({\bf A}) and ({\bf B}) with changing the pumping power. }\label{EmissionvsPower}
\end{figure}

\begin{figure}
\center
\includegraphics[scale=0.8]{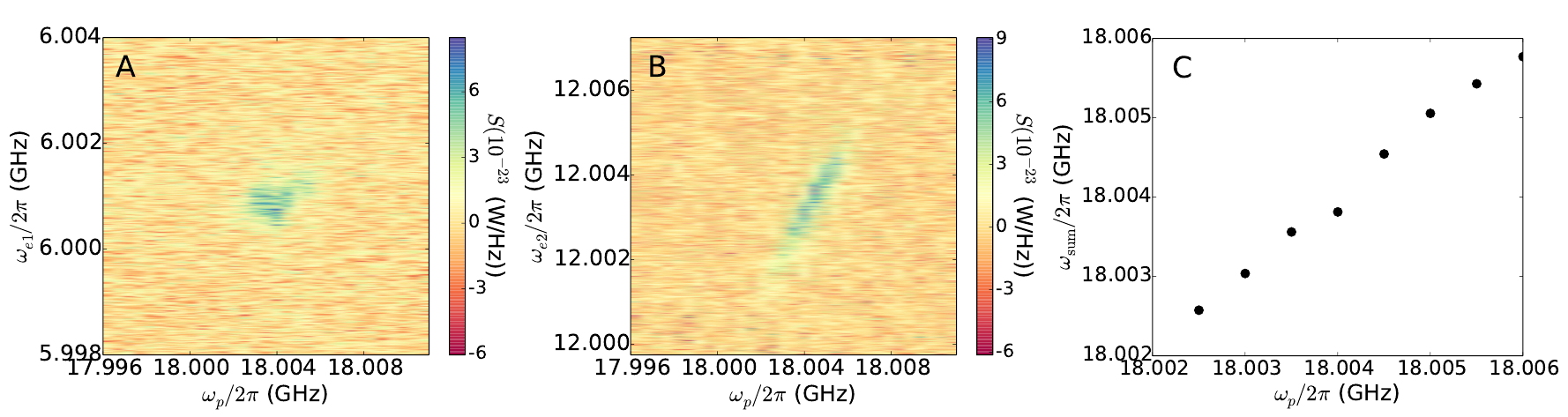}
\caption{Sum of the two center frequencies of emission spectra is equal to the pumping frequency with changing the pumping frequency. The emission spectra from the first mode ({\bf A}) and the second mode ({\bf B}) with changing the pumping frequency. ({\bf C}) The sum of the two center frequencies of emission spectra in ({\bf A}) and ({\bf B}) with changing the pumping frequency.}\label{EmissionvsFrequency}
\end{figure}

In the experiment, when we change the pumping power and pumping frequency, the sum of the two center frequencies of the emission spectra is always equal to the pumping frequency, as shown in Fig.~\ref{EmissionvsPower} and Fig.~\ref{EmissionvsFrequency}.

% ----------------------FIGURE -----------------------------------
%% ----------------------FIGURE -----------------------------------
% ----------------------FIGURE -----------------------------------
%% ----------------------FIGURE -----------------------------------

\subsection{Experimental setup for measuring the quenching of the phase diffusion and the correlation of the two-mode CEL}

\begin{figure}
\center
\includegraphics[scale=0.7]{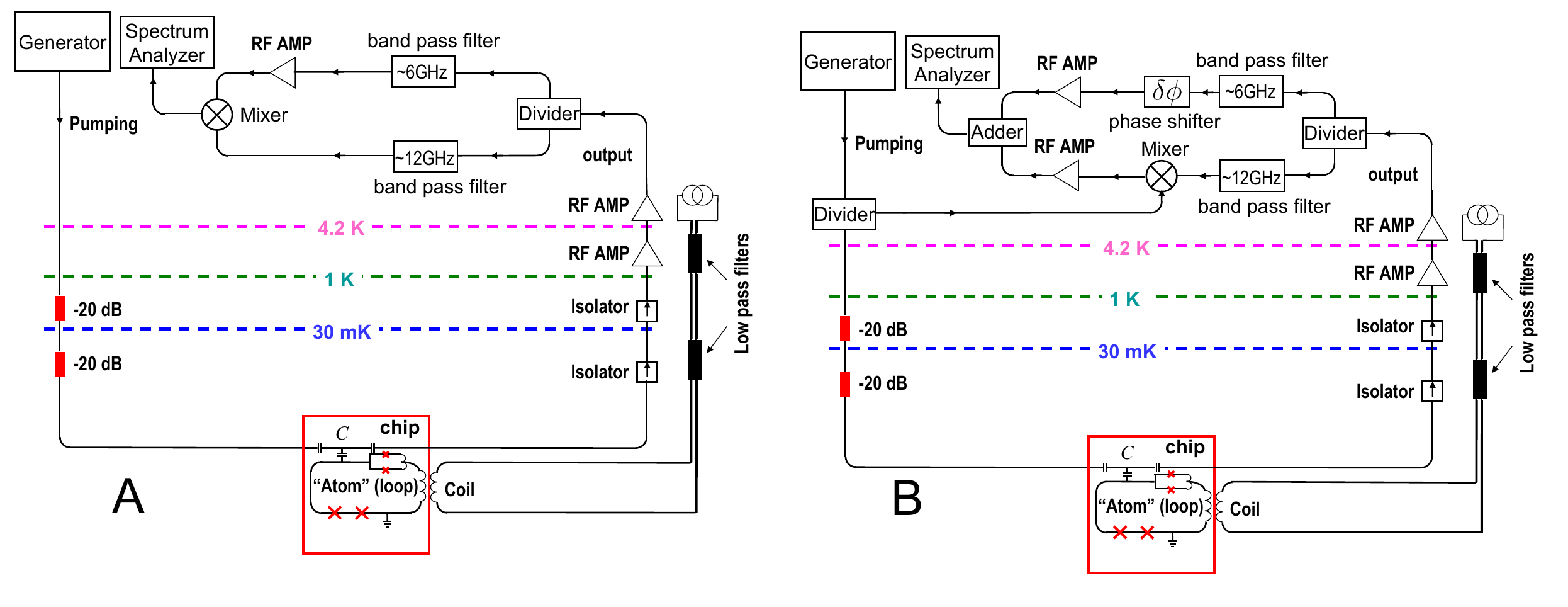}
\caption{({\bf A}) Experimental setup for measuring the quenching of the phase-diffusion noise of the two-mode CEL. ({\bf B}) Experimental setup for measuring correlation of the two-mode CEL.}\label{CorrelationSetup}
\end{figure}

The experimental setup used to demonstrate the quenching of the phase-diffusion noise of the two-mode CEL is presented in Fig.~\ref{CorrelationSetup} A.

The experimental setup used to demonstrate the correlation of the lasing fields with two diffent colors in the two-mode CEL is presented in Fig.~\ref{CorrelationSetup} B. It is necessary to seperately balance the gains in the two different chains to obtain maximal modulation in the interference fringes.
% ----------------------FIGURE -----------------------------------
%% ----------------------FIGURE -----------------------------------